\documentclass[namedreferences]{solarphysics}

\usepackage[hyperref,optionalrh]{spr-sola-addons} 
\usepackage{graphicx}        
\usepackage{color}           
\usepackage{breakurl}        

\renewcommand{\vec}[1]{{\mathbfit #1}}

\chardef\us=`\_

\begin{document}

\begin{article}
\begin{opening}

\title{Origin and Ion Charge State Evolution of Solar Wind Transients during 4 -- 7 August 2011.}

\author[addressref={aff1},corref,email={rodkindg@gmail.com}]{\inits{D.}\fnm{D.}~\lnm{Rodkin}\orcid{http://orcid.org/0000-0002-5874-4737}}
\author[addressref=aff1,email={goryaev\underline{ }farid@mail.ru}]{\inits{F.}\fnm{F.}~\lnm{Goryaev}\orcid{orcid.org/0000-0001-9257-4850}}
\author[addressref=aff2,email={pp25@st-andrews.ac.uk}]{\inits{P.}\fnm{P.}~\lnm{Pagano}\orcid{orcid.org/0000-0001-5274-515X}}
\author[addressref=aff3,email={gordon.gibb@ed.ac.uk}]{\inits{G.}\fnm{G.}~\lnm{Gibb}}
\author[addressref=aff1,email={slem@sci.lebedev.ru}]{\inits{V.}\fnm{V.}~\lnm{Slemzin}\orcid{orcid.org/0000-0002-5634-3024}}
\author[addressref=aff4,email={jshugai@srd.sinp.msu.ru}]{\inits{Yu.}\fnm{Yu.}~\lnm{Shugay}\orcid{orcid.org/0000-0003-3278-6557}}
\author[addressref={aff4,aff5},email={veselov@dec1.sinp.msu.ru}]{\inits{I.}\fnm{I.}~\lnm{Veselovsky}\orcid{orcid.org/0000-0002-0228-0320}}
\author[addressref={aff2},email={dhm@st-andrews.ac.uk}]{\inits{D. H.}\fnm{D. H. }~\lnm{Mackay}\orcid{orcid.org/0000-0001-6065-8531}}

\address[id=aff1]{P.N. Lebedev Physical Institute of the Russian Academy of Sciences, 53 Leninskiy Prospekt, Moscow, 119991, Russia}
\address[id=aff2]{School of Mathematics and Statistics, University of St Andrews, North Haugh, St Andrews, Fife, Scotland, KY16 9SS, UK}
\address[id=aff3]{School of Mathematics and Statistics, University of St Andrews, North Haugh, St Andrews, Fife, Scotland, KY16 9SS, UK; now at Edinburgh Parallel Computing Centre}
\address[id=aff4]{Skobeltsyn Institute of Nuclear Physics, Lomonosov Moscow State University, Leninskie gory, GSP-1, Moscow, 119991, Russia}
\address[id=aff5]{Space Research Institute of the Russian Academy of Sciences, 53 Leninskiy Prospekt, Moscow, 119991, Russia}

\runningauthor{D.Rodkin \textit{et al.}}
\runningtitle{Solar origins of transients}

\begin{abstract}

We present study of the complex event consisting of several solar wind transients detected by \textit{Advanced Composition Explorer} (ACE) on 4 -- 7 August 2011, that caused a geomagnetic storm with Dst$=-$110 nT. The supposed coronal sources -- three flares and coronal mass ejections (CMEs) occurred on 2 -- 4 August 2011 in the active region (AR) 11261. To investigate the solar origin and formation of these transients we studied kinematic and thermodynamic properties of the expanding coronal structures using the \textit{Solar Dynamics Observatory/Atmospheric Imaging Assembly} (SDO/AIA) EUV images and the differential emission measure (DEM) diagnostics. The \textit{Helioseismic and Magnetic Imager} (HMI) magnetic field maps were used as the input data for the 3D magnetohydrodynamic (MHD) model to describe the flux rope ejection \citep{pagano13a}. We characterize the early phase of the flux rope ejection in the corona, where the usual three-component CME structure formed. The flux rope ejected with a speed of about 200~km~s$^{-1}$ to the height of 0.25~R$_{\odot}$. The kinematics of the modeled CME front agrees well with the \textit{Solar Terrestrial Relations Observatory} (STEREO) EUV measurements. Using the results of the plasma diagnostics and MHD modeling, we calculated the ion charge ratios of carbon and oxygen as well as the mean charge state of iron ions of the 2 August 2011 CME, taking into account the processes of heating, cooling, expansion, ionization, and recombination of the moving plasma in the corona up to the frozen-in region. We estimated a probable heating rate of the CME plasma in the low corona by matching the calculated ion composition parameters of the CME with those measured \textit{in situ} for the solar wind transients. We also consider the similarities and discrepancies between the results of the MHD simulation and the observations.

\end{abstract}
\end{opening}

\section{Introduction}
     \label{S-Introduction}

The key problem of space weather forecasting is the prediction of geoeffective transient solar wind streams that are capable of causing geomagnetic storms. One of the most geoeffective solar wind (SW) transients are interplanetary coronal mass ejections (ICMEs). They are considered as interplanetary manifestations of coronal mass ejections (CMEs) associated with solar activity eruptive processes. Observational criteria and properties of ICMEs and their specific types, as magnetic clouds at 1 AU, can be found in many publications (\textit{e.g.} \citealp{gosling90,bothmer1998,zurbuchen06,fwebb12}, and references therein).

For the earliest prediction of ICME arrival to Earth one needs to understand the key factors which determine the formation of the transient plasma flows in the corona and their propagation in the heliosphere. From an observational point of view, a typical ICME develops in the corona in four main stages: (1) eruption of plasma with the formation and expansion of a flux rope in the corona observable by EUV telescopes up to the distance of 1.3\,--\,1.7~R$_{\odot}$ from the solar centre, (2) appearance of a CME in the field of view of white light coronagraphs above the limb at the distances $>$~2~R$_{\odot}$, (3) propagation of the CME in the heliosphere visible with wide-field coronagraphs or heliospheric imagers, (4) appearance \textit{in situ} of the solar wind transient with signatures of ICME -- significant deviations of the main parameters ($V_{\rm p}$, $n_{\rm p}$, $T_{\rm p}$, and $B$, which are the proton velocity, density, and temperature, and the magnetic field, respectively) from the ambient values \citep{rich04,rich10}. The tracking of a CME from the low corona to Earth using a qualitative morphological analysis of the pre-eruption structure and the subsequent events was presented by \citet{deforest2013}.

CMEs are the clearest evidence of ICME initiations. The source regions of CMEs in the solar corona can be identified and localized by their characteristic signatures: solar flares, flows, magnetic reconfiguration, EUV waves, jets, coronal dimmings or brightenings, filament eruptions, post-flare loop arcades revealed by continuous monitoring of the solar corona by EUV imaging using well-developed methods of data processing. \citet{cremades2004} have studied 276 CMEs in the period 1997--2002 and concluded that the topology of these CMEs and their orientation were defined by the location and magnetic configuration of the sources, in particular, the position of the neutral line between two opposite magnetic polarities. This characteristic has been recognized in the solar event of 2 August 2011, which is considered below.

Prediction of ICME arrival to Earth environment depends on the CME kinematics in the heliosphere. Solar wind models describe this propagation. The simplest one, the Archimedean spiral (or ballistic) model (ASM), is based on the assumption that the solar wind streams have constant velocity during the whole passage from the Sun to Earth \citep{nolte73a,mcneice11}. The ASM model does not take into consideration the evolution of the CME velocity in the heliosphere, so it can be used only to have an approximate determination of the arrival time interval with an uncertainty of half a day. The improved drag-based model (DBM) assumes that the dynamics of CMEs is dominated by the magnetohydrodynamic (MHD) aerodynamic drag \citep{cargill96,vrsnak01,owens04,cargill04,vrsnak04,vrsnak10,vrsnak07,vrsnak08,borg09,lara09}, \textit{i.e.} that above a distance of 20~R$_{\odot}$ CMEs faster than the ambient solar wind are decelerated, whereas those slower -- are accelerated by the ambient flow \citep{gopalswamy00}. A validation analysis shows that the mean uncertainty in predicting the arrival time of ICMEs by the DBM goes from 12.9 hours, when the deviation of CMEs from the Sun\,--\,Earth line is larger than their half-width, to 6.8 hours when such deviation is smaller \citep{vrsnak13,shi15}. Prediction of the arrival time by the kinematical models is based on the knowledge of the initial CME velocity at the distance of  $\sim$~20~R$_{\odot}$, which is subject to projection effect depending on the angle between the CME propagation direction and the Sun\,--\,Earth line. \citet{colaninno2009} proposed a method to derive the propagation direction by comparing masses of the CME structure determined from the Thomson scattering intensities observed by two \textit{Solar Terrestrial Relations Observatory} (STEREO) spacecraft located in different angular positions. They found that the direction ambiguity becomes reasonably small ($\leq$~20$^0$) for angular separations between spacecraft $>$~50$^0$. However, this method does not take into consideration the CME size and distribution of masses in the 3D CME structure, so it gives the center-of-mass velocity rather than the velocity of the leading edge commonly used in prediction.

The aforementioned estimations concern the propagation of single CMEs. In some cases, two or more successive CMEs propagate in direction to Earth. As a result of interaction, these CMEs change their kinematic parameters. \citet{temmer2012} described the case of 1 August 2010, when interaction of two CMEs resulted in strong deceleration of the overtaking second CME followed by their merging and further propagation as a single structure. The authors succeeded to simulate the kinematical evolution of the second CME using the drag-based model \citep{vrsnak13} varying the drag parameter $\Gamma$ value and the ambient solar wind speed from the region of interaction ($\sim$~35~R$_{\odot}$) to Earth. Other cases of interaction between CMEs and analysis of their kinematics by multipoint observations were described in \citet{mostl2012}, \citet{lugaz2012}, \citet{colaninno2015}, and in references therein.

Currently several sophisticated physics-based models exist for solar wind forecasting near Earth and beyond with the use of the MHD approach: the Wang-Sheeley-Arge (WSA)-Enlil model, MHD-Around-a-Sphere (MAS)-Enlil model, Space Weather Modeling Framework (SWMF), and their combinations \citep{jian15}. These models structurally consist of two main parts: the solar coronal and heliospheric components. The coronal part of the WSA, MAS, and SWMF models approximates the outflow at the inner boundary of the heliosphere based on synoptic magnetograms constructed from daily full-disk photospheric magnetograms \citep{arge00} using a semi-empirical model based on the potential magnetic field approximation. Then, the boundary outflow and magnetic field distributions are used as the initial data in MHD simulations, which describe the radial expansion and evolution of the solar wind plasma in the heliosphere.

The Enlil model is a 3D time-dependent heliospheric model based on ideal MHD equations \citep{odstril94,odstril96,odstril99}. The inner boundary of the Enlil model is placed at 21.5 or 30~R$_{\odot}$ beyond the outermost critical point and the outer boundary is taken from 2 to 5 or 10 AU. To predict the propagation of CMEs, the WSA-Enlil cone model additionally uses white-light images from \textit{Large Angle and Spectrometric Coronagraph} (LASCO: \citet{brueckner95}) and coronagraphs onboard STEREO \citep{howard08}. Such a model characterizes the basic properties of the CME, including velocity and size \citep{pizzo11,mays15}. This model focuses mainly on the prediction of the arrival time of the CME without localization of its coronal origin. \citet{mays15} in their statistical analysis of 35 CMEs between January 2013 and July 2014 concluded that the mean estimated error of the CME arrival by the Enlil model was 12.3 hours. Inaccuracy of current models to a considerable degree arises from insufficient knowledge of the interaction of the CME plasma with the ambient solar wind and other wind components in the heliosphere. Information about these processes can be obtained from the fast progressing heliospheric tomography based on the interplanetary scintillation (IPS) in radio waves \citep{manoh10}.

Among other parameters of solar wind transients, the ion charge composition of the plasma is one of the important identifiers of their origin, whereas it depends on the parameters of the source and remains practically unchanged during the solar wind propagation in the heliosphere. Analysis of the ion charge state and mass composition of the solar wind plasma helps to separate its different components and to determine its source \citep{fisk98,zhao09,zhao14,kilpua14,wang12}.

It is believed that the ion charge state of the solar wind, registered at Earth$^\prime$s orbit, approximately corresponds to its state in the corona at the altitude where it is "frozen-in" (a transition to the limiting case, where the ionization and recombination times of the plasma in the corona are of the order or larger that the time of the solar wind propagation in the heliosphere \citep{hundhausen68}). The frozen-in condition is valid at distances between 1.5 to 4~R$_{\odot}$, which depends on the type of ion and the level of activity \citep{feldman05}.

The main factors that determine the ion composition of the CME plasma and its evolution in the corona are temperature, density, and mass velocity, which depend on the level of solar activity. At higher activity, plasma temperature and density in the source and average CME speed increase. During the expansion of the CME plasma from the origin to the frozen-in region, its ion composition evolves, due to various processes, such as heating by an energy release from the flare site and cooling by heat conduction, radiation losses and adiabatic expansion. From \textit{in situ} measurements it was established that the faster CME and ICMEs, as a rule, have higher ion charge states in comparison to the slower ones \citep{Gopals13}.

\citet{Gruesbeck11} presented a procedure for deriving the ion composition of CMEs in the corona. To achieve this, they used \textit{in situ} measurements of the ion charge states of C, O, Si, and Fe and interpreted them in the context of a model for the early evolution of the CME plasma. They obtained, in particular, a best fit for the data provided by an initial heating of the plasma, followed by cooling expansion. \citet{lepri12} presented an analysis and comparison of the heavy ion composition, observed during the passage of an ICME at the \textit{Advanced Composition Explorer} (ACE) and \textit{Ulysses}. They compared the ion composition, obtained across the two different observation cuts through the ICME, with predictions for heating during the eruption, based on models of the time-dependent ionization balance throughout the event. The authors of both papers based their considerations only on assumptions about conditions in the coronal sources without comparing them with measurements.

\citet{Lynch11} and \citet{Reinard12} used large scale non-ideal 2.5D MHD simulations of the solar wind and investigated the ion charge state composition during the CME propagation. They found only a qualitative matching between the observation and model in the charge state enhancements in the flux rope material and in its front, because a 2.5D model is not suitable for the proper quantitative description of the ionization state of plasma elements. The 3D MHD model presented in this article \citep{pagano13a} certainly addresses these improvements, however the very different spatial domains of the two studies make difficult a direct comparison. \citet{Pagano08} had already sucessfully derived the ion composition from the post-processing of MHD simulations. They explained several observational features of shocks connected to CMEs by reconstructing the out-of-equilibirium ionization state of O~\textsc{\MakeLowercase{VI}} and Si~\textsc{\MakeLowercase{XII}}, when an MHD shock undertakes the plasma.

In this article, we present a complex method to predict the ion composition of the solar wind transients using both numerical simulations and direct observations of the CME initiation in the corona. Such approach enables us to establish relations between parameters of the CME source in the corona and the resulting \textit{in situ} parameters of the solar wind transients. We consider a case of three solar events -- X-ray flares and CMEs that occurred on 2 -- 4 August 2011, which led to significant disturbances of the solar wind near Earth on 4 -- 7 August 2011 and produced a strong geomagnetic storm with a minimum value of Dst$=-$110 nT \citep{yigit16}. In our work, we focus on the investigation of the solar part of this complex event under favorable conditions, when the associated coronal phenomena had arisen near the solar disk center and the resulting CMEs were observed in quadrature by the \textit{Sun Earth Connection Coronal and Heliospheric Investigation} (SECCHI: \citealp{howard08}) instruments onboard STEREO-A and B spacecraft and LASCO C2 coronagraph onboard SOHO.

The paper consists of eight sections. Introduction gives a review of methods for identification of ICMEs and their solar sources. In Section 2, we present a general description of three eruptions on 4 -- 7 August 2011. Sections 3 -- 6 contain a detailed analysis of the first event of 2 August 2011: the observational data, diagnostics of the CME plasma, description of the 3D MHD model of the flux rope ejection, comparison of the results of numerical simulations with observations, and analysis of the plasma ion composition in the solar source and its evolution in the corona. The final sections summarize the results of our study. A detailed study of the second and third events of 3 and 4 August, when two CMEs interacted in the heliosphere, will be presented in a next article.

In the analysis, we used the data from the \textit{Solar Dynamic Observatory} (SDO): the solar EUV images from the \textit{Atmospheric Imaging Assembly} telescope (AIA: \citealp{lemen12}) and the photospheric magnetic field maps from the \textit{Heliospheric and Magnetic Imager} instrument (HMI: \citealp{schou12}). Diagnostics of flares in the coronal sources were fulfilled using the \textit{Geostationary Operational Environmental Satellite} system (GOES) X-ray data, temperature and density of the outflow plasma were defined by the differential emission measure (DEM) method using the AIA multi-wavelength EUV images. The solar wind data including the ICME parameters and charge states of C, O and Fe ions were extracted from the ACE data \citep{stone98}.

\section{The Solar Wind Transients of 4 -- 7 August 2011 and Their Solar Sources}
\label{S-ICMEs}
\begin{figure}    
\centerline{
\includegraphics[width=1.0\textwidth,clip=]{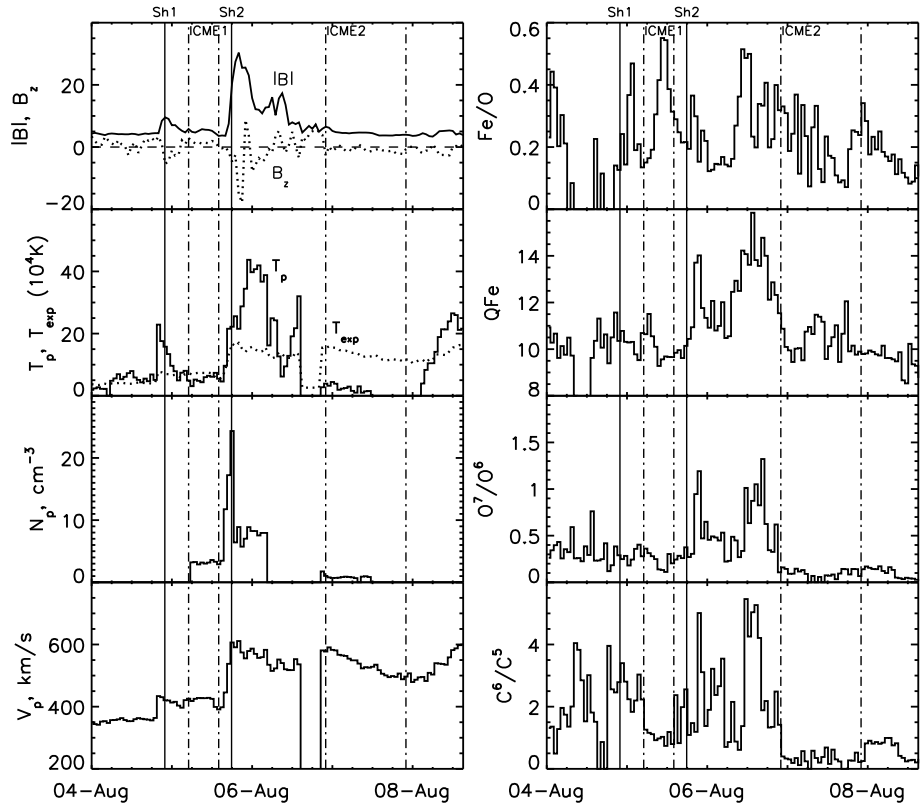}
}
\caption{ACE solar wind data for 4 -- 8 August 2011. The times of shocks, Sh1 and Sh2, that correspond to geomagnetic storm sudden commencements \citep{rich10} are marked by the solid lines. The start and end of ICME1 and ICME2 are marked by the dot-dashed lines.
}
\label{acedata}
\end{figure}

Figure \ref{acedata} shows the level 2 one-hour averaged ACE solar wind data for the period 4 -- 8 August 2011. The solar wind transients identified in these data were referenced in three International Study of Earth-Affecting Solar Transients (ISEST)\footnote{http://solar.gmu.edu/heliophysics/index.php/ISEST} related databases. The ICME and CME lists of George Mason University (GMU)\footnote{http://solar.gmu.edu/heliophysics/index.php/GMU\_CME/ICME\_List} and University of Science and Technology of China (USTC)\footnote{http://space.ustc.edu.cn/dreams/wind\_icmes} describe only one ICME of 6 August 2011 classified as the EJ+CIR+SH (ejecta + corrotating interaction region + shock) event (GMU list) or SH+EJ (shock + ejecta) event (USTC list), which led to the geomagnetic storm of 6 August, 12:00 UT, with Dst$=-$110 nT. The GMU list associates ICME with the M9 flare and halo CME that occurred on 4 August 2011. The Richardson and Cane ICME list (hereafter RC list)\footnote{http://www.srl.caltech.edu/ACE/ASC/DATA/level3/icmetable2.htm} mentions two shocks and two ICMEs, indicated in Table~\ref{T-ICMEs} and marked in Figure \ref{acedata}. The shock times in the RC list correspond to the geomagnetic storms sudden commencements  that accompany the shocks reaching the Earth$^\prime$s magnetosphere \citep{rich10}. The first ICME showed small enhancements of magnetic field, proton density, and speed above the background level. The temperature-related ion ratios, C$^{6+}$/C$^{5+}$, O$^{7+}$/O$^{6+}$, and the mean charge of iron ions, Q$_{\mathrm{Fe}}$, slightly exceeded the background. Only the structure-related Fe/O ratio showed a noticeable increase to the value of 0.55, which corresponds to a first ionization potential (FIP) bias of $\sim$~3  and  evidences a presence of plasma from closed magnetic structures in the source region. The second ICME displayed a jump of the proton speed to 610~km~s$^{-1}$ with moderate enhancements of the magnetic field magnitude and proton density. Thus, the typical ICME signatures \citep{zurbuchen06,rich04}, except the decreased proton temperature and the ion composition, were slightly shown by both ICMEs. However, identification of the second ICME in the RC list is rather ambiguous because other signatures such as enhanced magnetic field and ion charge state did not appear. The negative values of the $z$ component, $B_{\rm z}$, of the interplanetary magnetic field (IMF) ($-$5 and $-$17~nT in the geocentric solar magnetospheric (GSM) coordinate system) were registered in the sheaths that followed the shocks, being the most likely cause of the geomagnetic storm of 6 August 2011. The development of this storm was considered in detail by \citet{yigit16}.

\begin{figure}    
\centerline{
\includegraphics[width=0.5\textwidth,clip=]{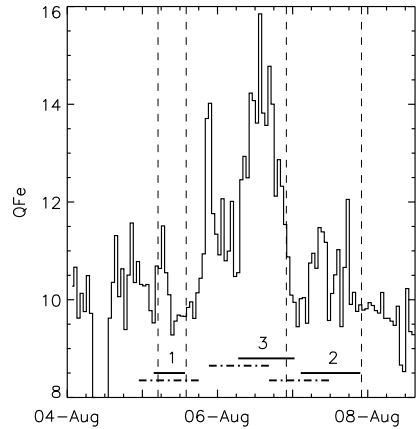}
}
\caption{The ACE Q$_{\mathrm{Fe}}$ data, the solar wind transients identified by the Q$_{\mathrm{Fe}}$ enhancements (solid lines with numbers) and their arrival time slots given by the Advanced Drag Model (the dash-dotted lines). The numbers designate the association with the CMEs listed in Table 2
}
\label{ionstate}
\end{figure}

An application of the ballistic propagation model (assuming that the ICME speed between the Sun and Earth is equal to its \textit{in situ} value taken from the RC list) gives a preliminary time period for the solar events that probably produced these ICMEs from 31 July 2011, 18:00 UT to 4 August 2011, 13:00 UT. During this period, three flares and three CMEs, directed to Earth, occurred in active region (AR) 11261 (Table~\ref{T-Flares}). In order to identify the solar sources of the solar wind transients, we used simulations of two heliospheric CME propagation models: the Advanced Drag Model(ADM) \citep{vrsnak13} and the WSA-Enlil Model \citep{pizzo11}. For the three CMEs, listed in Table 2, taking an asymptotic solar wind speed of 350~km~s$^{-1}$ the ADM gives transit times of 74.28, 84.48, and 51.08 hours with a mean uncertainty of $\sim$~10 hours \citep{shi15}. These transit times correspond to the time slots of the ICME arrival time (to Earth): 4 August 2011, 22:50 UT -- 5 August 2011, 18:50 UT, 6 August 2011, 16:30 UT -- 7 August 2011, 12:30 UT, and 5 August 2011, 21:20 UT -- 6 August 2011, 17:20 UT (Figure~\ref{ionstate}). The slots associated with CME1 and CME2 agree well in time with the ICMEs in the RC list and enhancements in the ion charge state composition derived from ACE data. The slot associated with the fast CME3 coincides with the noticeable enhancements in the ion charge state and magnetic field magnitude, but was not identified as an ICME. It should be noted that the ADM time slots coincide with the ion charge state transients with a mean discrepancy of 5.6 hours. The minimum, maximum and mean values of the ion composition parameters for these transients are given in Table~\ref{T-ionstate}.

\begin{figure}    
\centerline{
\includegraphics[width=0.7\textwidth,viewport=0 0 250 227,clip=]{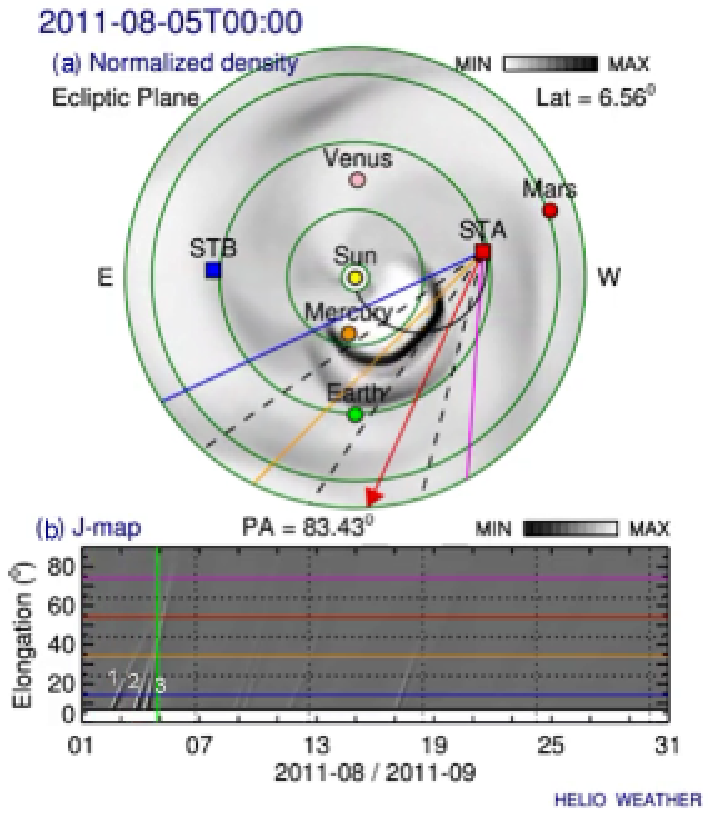}
}
\caption{(a) The normalized density map in the ecliptic plane and (b) the J-map for the three solar wind transients obtained with the WSA-Enlil model on 5 August 2011, 00:00 UT (http://helioweather.net/archive/2011/08/)}

\label{enlil}
\end{figure}

To understand the nature of the third transient, we consider the results of the simulation of this complex event by the WSA-Enlil model, presented in the Enlil solar wind prediction helioweather database\footnote{http://helioweather.net/archive/2011/08/}. It was found that CME3 (started on 4 August 2011), due to its higher speed, it overran CME2, which started on 3 August 2011, at the distance of $\sim$~0.6 AU, when CME1 had already reached Earth. In Figure~\ref{enlil} we present the map of the normalized plasma density in the ecliptic plane for 5 August 2011, 00:00 UT, and the J-map for the three solar wind transients from the helioweather database. As a result of interaction between the second and third CMEs, the merged cloud reached the Earth on 6 -- 8 August 2011, producing the observed variation of the ion composition.

In Section 3 we consider in general the complex solar wind event of 4 -- 7 August 2011, its coronal sources, and in detail the formation of CME1 that originated on 2 August 2011. The features of the second and third CMEs and their relations with the solar origins will be studied in a following publication.

\begin{table}  
\caption{Data of the ICMEs on 4 -- 5 August 2011 presented in the Richardson and Cane ICME list (http://www.srl.caltech.edu/ACE/ASC/DATA/level3/icmetable2.htm)}

\label{T-ICMEs}
\begin{tabular}{ccccc}     
  \hline                   
N  & Shock & ICME start & ICME end & V$_{\mathrm{max}}$ \\
  of event   & date/time, [UT]     & date/time, [UT] & date/time, [UT] & [km~s$^{-1}$] \\
     \hline
1 & 4 Aug. 2011  & 5 Aug. 2011  & 5 Aug. 2011  & 440 \\
  & 21:53 & 05:00 & 14:00 & \\
2 & 5 Aug. 2011  & 6 Aug. 2011  & 7 Aug. 2011  & 610 \\
  & 17:51 & 22:00 & 22:00 & \\
  \hline
\end{tabular}
\end{table}

\begin{table}  
\caption{Flares and CMEs occurred on 2 -- 4 August 2011 using data from GOES, STEREO-A/COR2 (A) and SOHO/LASCO (L)}
\label{T-Flares}
\begin{tabular}{ccccccc}     
  \hline                   
Date & Flare onset & Flare & CME (A)  & V$_{\mathrm{CME (A)}}$ & CME (L) & V$_{\mathrm{CME (L)}}$ \\
of event & time, [UT] & class & onset time, [UT] & [km~s$^{-1}$] & onset time, [UT] & [km~s$^{-1}$] \\
     \hline
2 Aug. 2011 & 05:19 & M1.4 & 05:54 & 781 & 06:36 & 712 \\
(CME1) & & & & & & \\
3 Aug. 2011 & 13:24 & M6.0 & 13:54 & 892 & 14:00 & 610 \\
(CME2) & & & & & & \\
4 Aug. 2011 & 03:41 & M9.3 & 04:12 & 1193 & 04:12 & 1315 \\
(CME3) & & & & & & \\
  \hline
\end{tabular}
\end{table}

\begin{table}   
\caption{Ion composition parameters of the solar wind transients determined from the QFe enhancements (Figure 2)
}
\label{T-ionstate}
\begin{tabular}{cccccc}     
  \hline                   
N$_{\mathrm{trans}}$ & start/end, & C$^{6+}$/C$^{5+}$ & O$^{7+}$/O$^{6+}$ & Q$_{\mathrm{Fe}}$ & Fe/O \\
& [UT] & min/max/mean & min/max/mean & min/max/mean & min/max/mean \\
     \hline
1 & Aug. 5  03:00/ & 0.74/3.03/1.31  & 0.11/0.40/0.25 & 9.28/11.5/10.1 & 0.13/0.55/0.31 \\
  & Aug. 5  14:00 &  &  &  &    \\
2 & Aug. 7  02:00/ & 0.20/0.69/0.36 & 0.03/0.15/0.10 & 9.45/12.05/10.45 & 0.07/0.37/0.19 \\
  & Aug. 7  22:00 &  &  &  &   \\
3 & Aug. 6  06:00/ & 0.31/5.46/2.20 & 0.09/1.32/0.52 & 9.94/15.85/12.74 & 0.14/0.51/0.31 \\
  & Aug. 7  01:00 &  &  &  &   \\
\hline
\end{tabular}
\end{table}

\section{Formation of the CME of 2 August  2011 in the Corona}
\label{S-2aug}

\subsection{Kinematics} 
  \label{S-kinematics}
We studied the formation of three CMEs on 2 -- 4 August 2011 in the low corona using the SDO/AIA images in different wavelength channels. After preliminary processing level 1 to level 1.5 data we produced running-difference images to identify the moving coronal structures associated with a CME. These structures are seen as expanding loops, but in fact they represent projections of the expanding erupting shells integrated over their legs along the line-of-sight (LOS). In the studied events the largest contrast was seen in the 211{\AA} channel images.

\begin{figure} 

\centerline{\hspace*{0.005\textwidth}
\includegraphics[width=0.5\textwidth,clip=,bb=0 10 550 545]{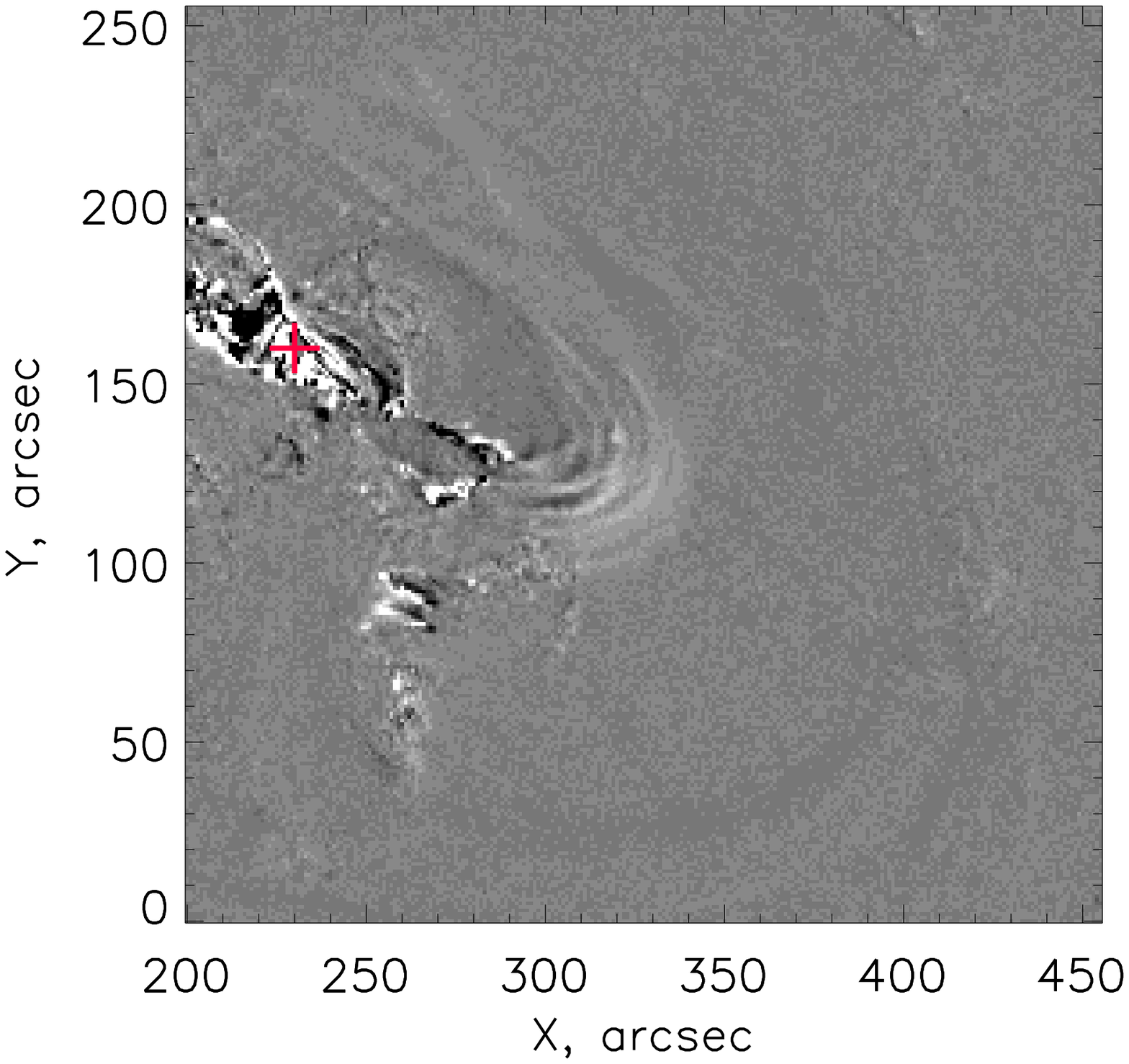}
\hspace*{-0.020\textwidth}
\includegraphics[width=0.5\textwidth,clip=,bb=0 10 550 545]{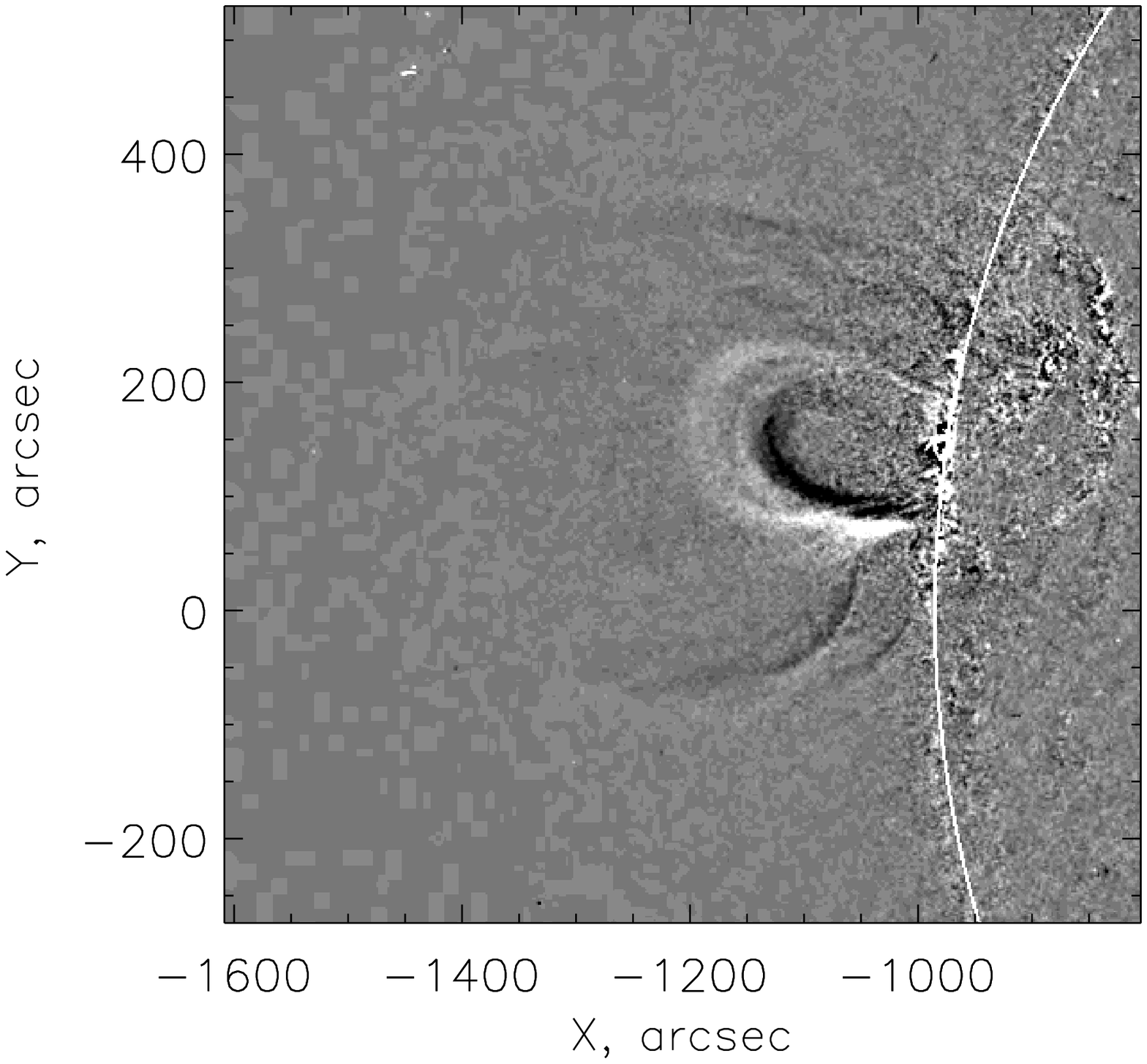}
}
\vspace{-0.45\textwidth}   
\centerline{\Large \bf     
\hspace{0.15 \textwidth}  \color{white}{(a)}
\hspace{0.39 \textwidth}  \color{white}{(b)}
\hfill}
\vspace{0.40\textwidth}    
\centerline{\hspace*{0.025\textwidth}
\includegraphics[width=0.5\textwidth,clip=,bb=10 20 560 545]{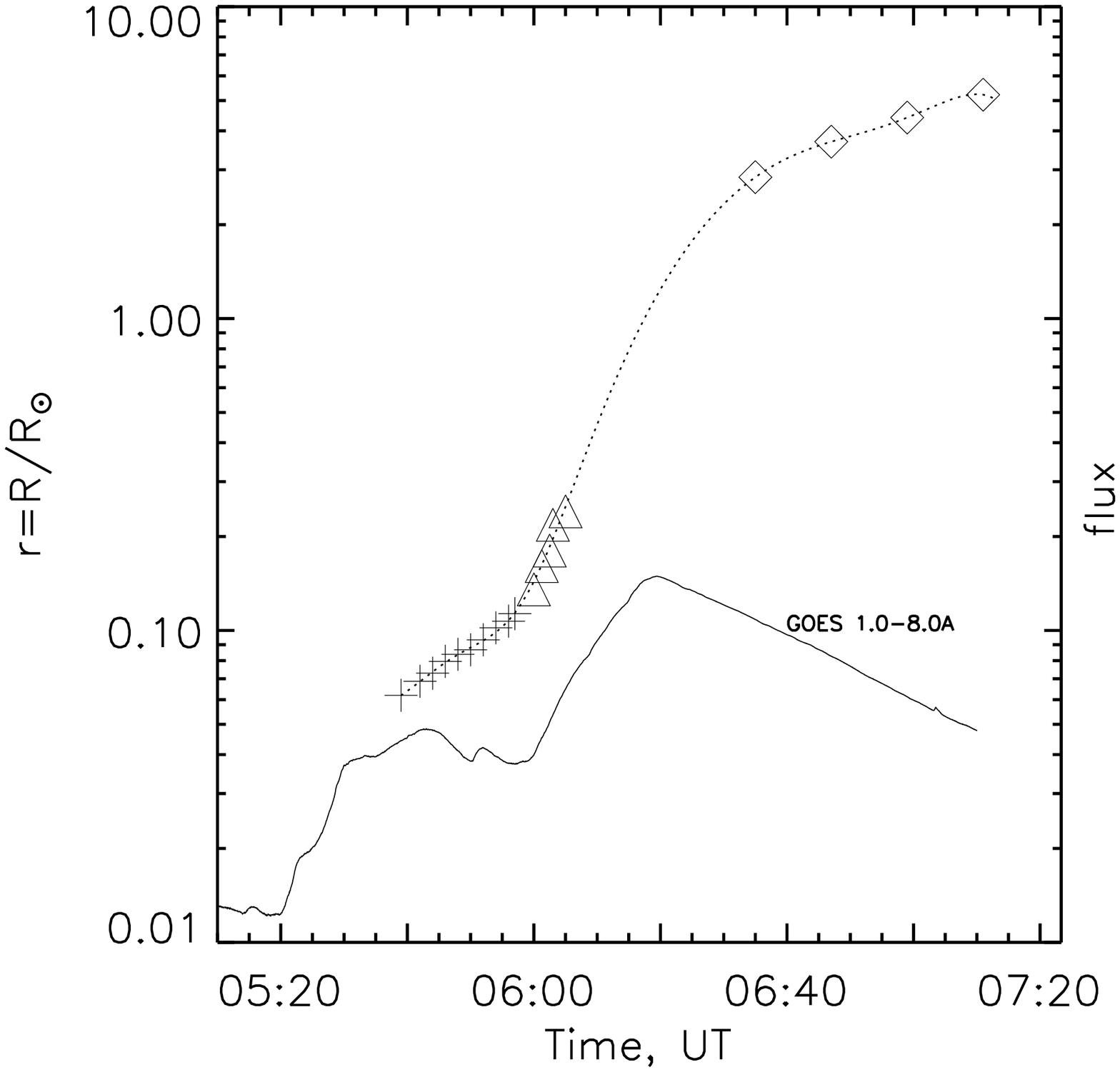}
\hspace*{-0.015\textwidth}
\includegraphics[width=0.5\textwidth,clip=,bb=15 20 560 545]{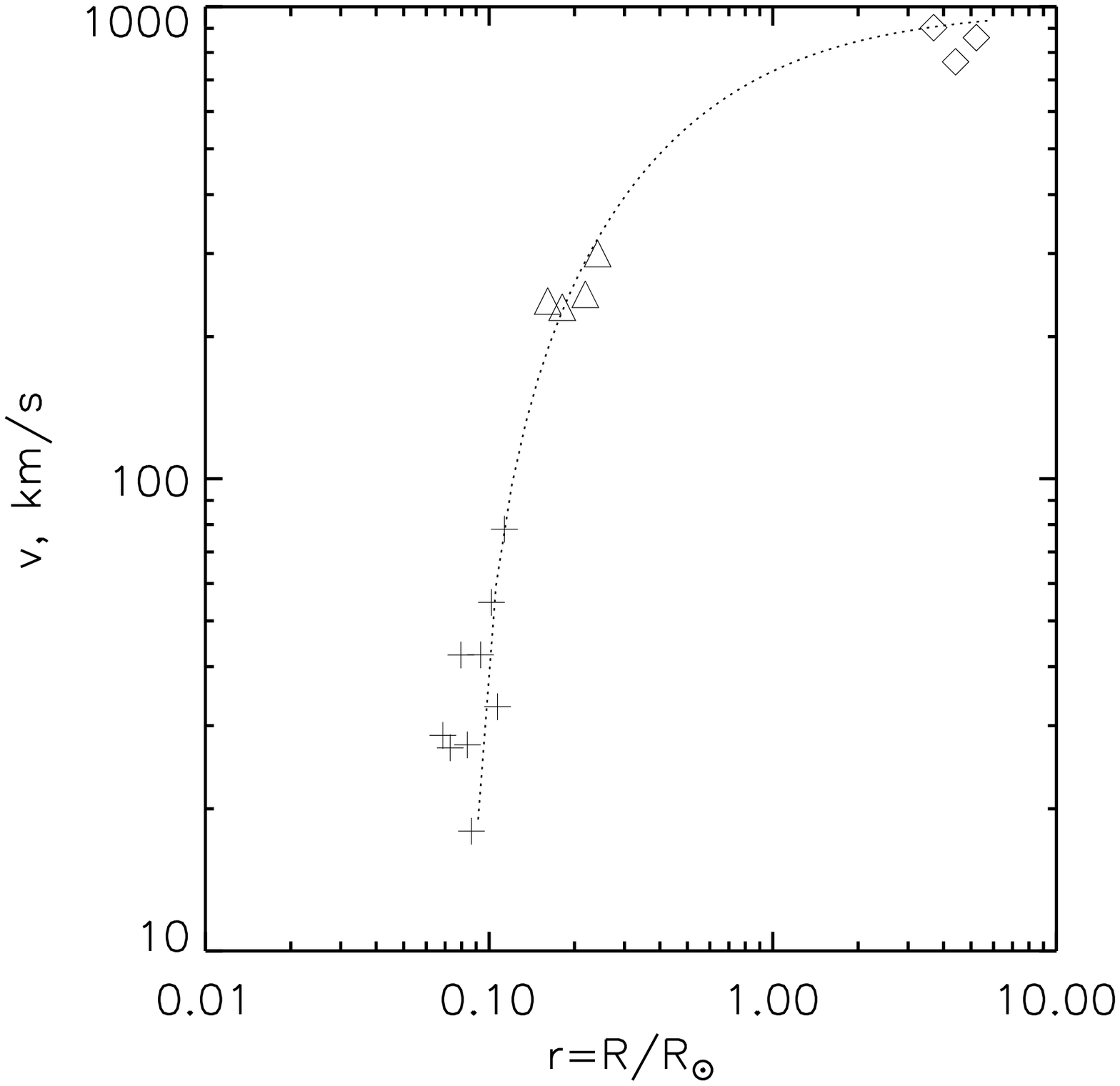}
}
\vspace{-0.45\textwidth}
\centerline{\Large \bf     
\hspace{0.15 \textwidth} \color{black}{(c)}
\hspace{0.39 \textwidth}  \color{black}{(d)}
\hfill}
\vspace{0.45\textwidth}    
\caption{Formation of the CME on 2 August 2011. (a) The running difference image (AIA, 211~\AA) shows the erupted structures in projection on the disk at 05:58:02 UT. (b) The running-difference image of the CME seen by STEREO-A/EUVI in 171~{\AA} at 06:02:15 UT. (c) The dotted line shows the dependence of the CME expansion height (the origin is marked by the cross in Figure 4a) on time. Crosses correspond to the brightest structure at the disk seen by AIA, triangles to the CME expansion above the limb seen by STEREO-A/EUVI, and the diamonds to the CME expansion seen by LASCO/C2. The solid line corresponds to the the flux from GOES 1.0 -- 8.0~{\AA}  in 10$^{-4}$~W~m$^{-2}$) (the flare occurred on 2 August 2011). (d) Dependence of the CME speed on time seen by AIA, EUVI and LASCO (symbols are the same as in (c)). The dotted line shows the fitting function (see text).
}
\label{rundif}
\end{figure}

Figure~\ref{rundif}a represents a group of expanding loop-like features in the 211{\AA} running-difference image at 05:58:02 UT, 2 August 2011, in the initial stage of CME1 formation (Table~\ref{T-ICMEs}). These structures apparently represent projections of the eruption shell on the disk plane. Nevertheless, they disappeared at $\sim$~06:00 UT, probably, due to the CME take-off or because of heating, as it will be shown in the Section 4. The distance-time graphs in Figure~\ref{rundif}c show the height variation of the CME in the low corona, seen in EUV AIA images (channel 211{\AA}), and above the limb, seen by STEREO-A/EUVI in 171{\AA} (Figure~\ref{rundif}b) and in the field of view of the LASCO C2 coronagraph. We associated the transverse distances from the LASCO data with height, assuming the self-similar expansion of the CME, when the vertical, $d_{\rm r}$,  and horizontal, $d_{\rm h}$,  displacements are in the same relation as the radial speed of the CME, $v_{\rm r}$, measured by STEREO-A/COR2, and the transverse speed, $v_{\rm h}$, measured by LASCO: $d_{\rm r} / d_{\rm h} \simeq v_{\rm r} / v_{\rm h}=$ 1.09. The acceleration phase of the CME corresponds to the second peak of the flare flux profile (bottom in Figure~\ref{rundif}c).

Figure~\ref{rundif}d shows the dependence of the CME speed on height, calculated from the data of different instruments.Projection effects are not important in our case. The WSA-ENLIL 3D simulations show that the leading edge of the CME structure in the ecliptic plane deviates from the Sun\,--\,Earth line westward no more than on 10$^0$, whereas the position angle of STEREO-A at that time was 100.5$^0$.  Thus, the data in Figure~\ref{rundif}d, obtained from observations by COR2 on STEREO-A, represent the real radial velocity of the CME. During the expansion, the CME speed increased from 26~km~s$^{-1}$ at $h_{\rm r}\simeq$~0.06~R$_{\odot}$ to $\sim$~800~km~s$^{-1}$ at 5~R$_{\odot}$.

\subsection{Plasma Diagnostics} 
  \label{S-diagnostics}
To derive the plasma parameters from the observed image data on 2 August 2011, we applied a plasma diagnostics methods for the flare region and the moving coronal structure associated with the CME at five different times. We used the differential emission measure (DEM) analysis to evaluate averaged electron temperatures and densities of the plasma structures under consideration. The DEM analysis was carried out, using intensities in six SDO/AIA EUV channels: 94~{\AA}, 131~{\AA}, 171~{\AA}, 193~{\AA}, 211~{\AA}, and 335~{\AA}. In all five positions along the direction of propagation we built the light curves of the mean intensities in 4$\times$4 arcsec boxes as a function of time and determined the fluxes in each spectral channel as the maximal value minus the mean background. The background corresponds to the quiet corona before the moving CME structure reaches it. The intensity flux $F_i$ in the channel {\it i} can be written as

   \begin{equation}
   \label{Eff_temp}
     F_i = \int\limits_{\Delta T} G_i(T) DEM(T) dT \, ,
   \end{equation}
where $G_i(T)$ is the temperature response function of the passband {\it i}, and $DEM(T)$ is the DEM distribution of the plasma.
To retrieve the function $DEM(T)$, we have used a DEM technique, based on the probabilistic approach to the inverse problem (see, \textit{e.g.} \citealp{urnov07,goryaev10,urnov12} for the details). A DEM temperature distribution for each plasma structure under consideration was then used to determine an effective temperature, $T_{\mathrm{eff}}$, according to the formula

   \begin{equation}  \label{Eff_temp}
     T_{\mathrm{eff}} = \frac{\int\limits_{\Delta T} T G(T) DEM(T) dT}{\int\limits_{\Delta T} G(T) DEM(T) dT} \, ,
   \end{equation}
where $G(T)=\sum_i G_i(T)$ is the total temperature response for all channels. The averaged electron density in a given plasma structure was then estimated, using the total emission measure (EM) and the plausible geometry of the corresponding plasma structure.

\begin{figure}    
\centering
\includegraphics[scale=0.6]{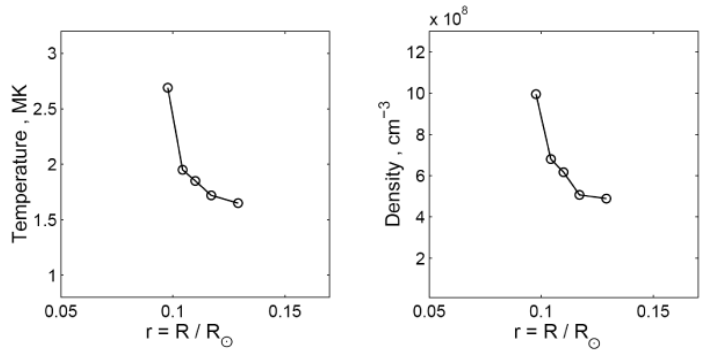}
\caption{Electron temperature and density evolution for the expanding plasma structure determined by the DEM method from the AIA images on 2 August 2011.}
\label{tem_den_evol}
\end{figure}

The averaged temperatures and densities for the flare on 2 August 2011 are given in Table \ref{T-Diagn-Flares}. Furthermore, the plasma parameters for the moving CME structure on 2 August 2011 in the range 0.1~--~0.15~R$_{\odot}$ from the solar surface are presented in Figure \ref{tem_den_evol}. It is worth noting that the temperatures in the moving plasma structure are noticeably lower than in the corresponding flare.

\begin{table}   
\caption{Averaged values of the electron temperature and density in the flare on 2 August 2011 (the parameters were obtained in the region marked with a red cross on the Figure 4a).}
\label{T-Diagn-Flares}
\begin{tabular}{ccc}     
  \hline
Time, [UT] & $T_{\mathrm{eff}}$ ,  [MK] & $n_e$ , [cm$^{-3}$]  \\
     \hline
 05:32 & 9.12 & $1.53\times10^{9}$  \\
 05:44 & 7.31 & $2.48\times10^{9}$  \\
 05:50 & 7.19 & $2.40\times10^{9}$  \\
  \hline
\end{tabular}
\end{table}

\section{Numerical Simulation of the Flux Rope Ejection} 
      \label{S-features}

To model this specific flux rope ejection in the solar corona,
we use an ideal 3D MHD simulation, coupled with a continuous time series
evolution of the magnetic field through a series
of quasi-static non-linear force-free (NLFF) fields .
The latter technique is used to produce a non-potential initial condition
that is used in the ideal MHD simulation.
This approach is a combination of the models, presented in
\citet{pagano13a} and \citet{gibb14},
that has been specifically tuned for the present simulation.

The key idea is to describe the flux rope formation and the conditions prior to the onset of the ejection
with the time series of NLFF fields, suited
for a slow, quasi-static, and magnetically dominated evolution,
and to adopt a full MHD description
for when the evolution of the system becomes
fast, out of equilibrium and in a multi-$\beta$ regime, where $\beta$ is the rate of the plasma pressure to the magnetic pressure.

\subsection{NLFF Time Series} 
  \label{S-equations}
In order to describe the slow evolution of the region of interest,
a continuous time series of NLFF fields are generated
from the corresponding time series of magnetograms \citep{mackay11,gibb14}.
In the present application we use 50 magnetograms
from 31 July 2011 at 05:00:41 UT to 2 August 2011 at 06:00:41 UT sampled every 60 minutes.
The procedure and set of equations solved is the same as in \citet{mackay11}.

The time series of NLFF fields is constructed assuming four closed boundaries at the sides
for the 3D box.
The bottom boundary, representing the solar surface, is forced to have magnetic flux balance
and the top boundary is set to be open.

\begin{figure} 
\vspace*{0.1\textwidth}
\centerline{\hspace*{-0.05\textwidth}
\hspace{0.07\textwidth}
\includegraphics[scale=0.17,viewport=0 0 700 550,clip=]{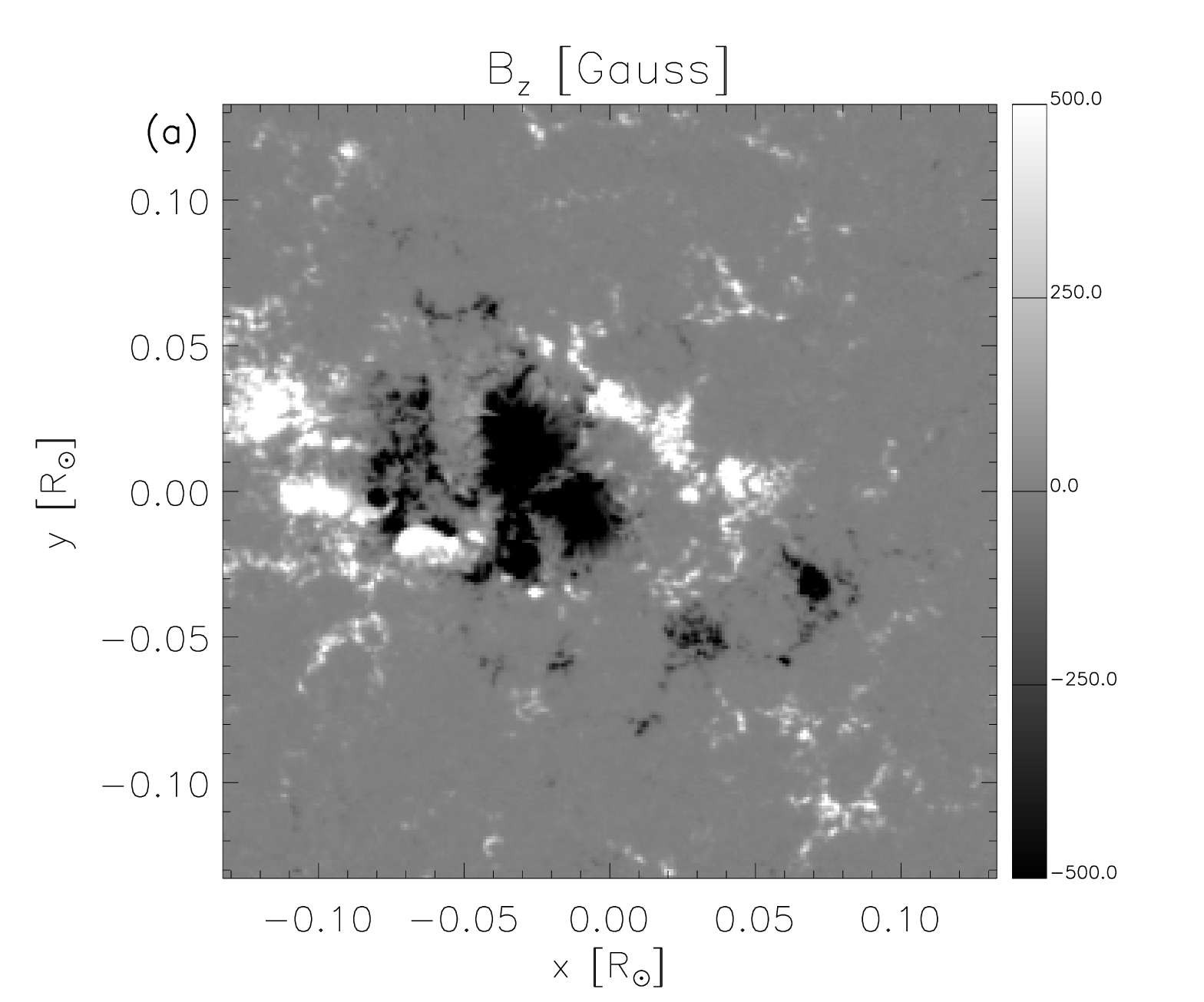}
\hspace{-0.01\textwidth}
\includegraphics[scale=0.35,viewport=55 0 350 250,clip=]{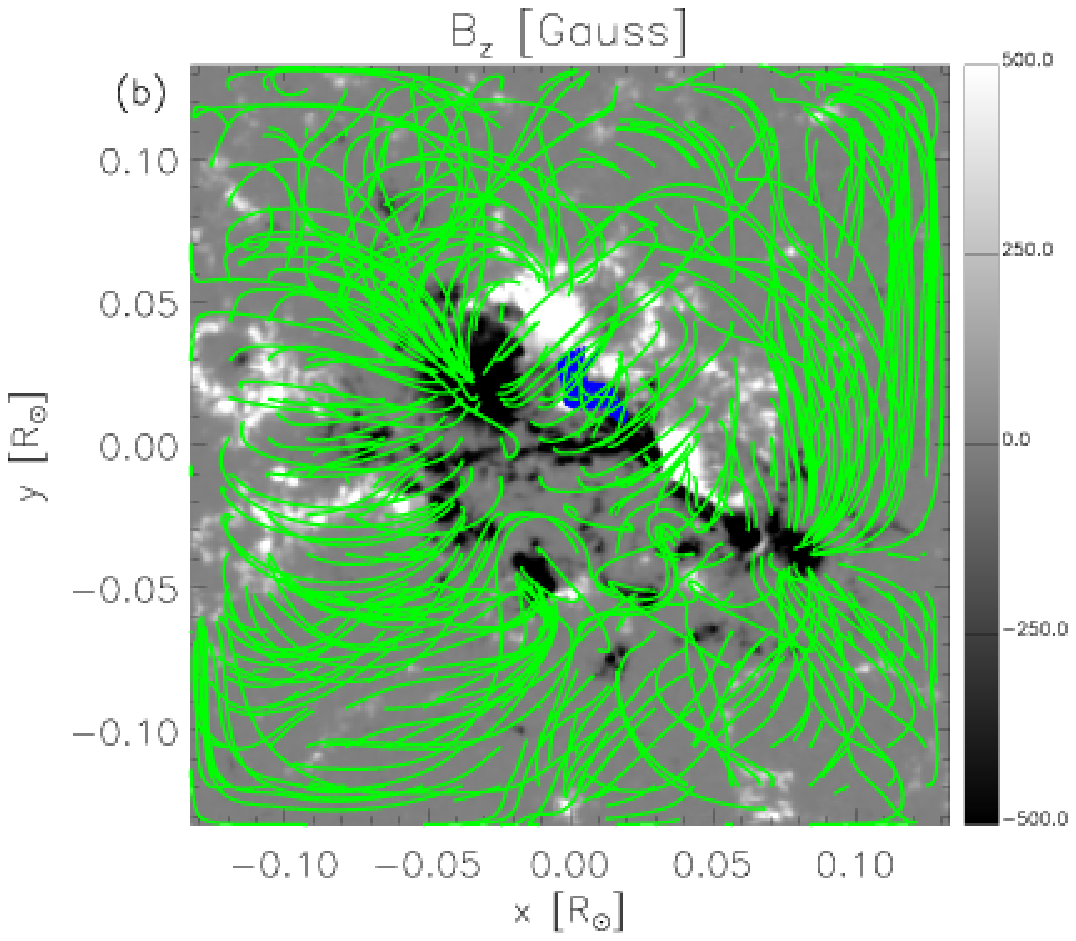}
\hspace{-0.01\textwidth}
\includegraphics[scale=0.17,viewport=55 0 700 550,clip=]{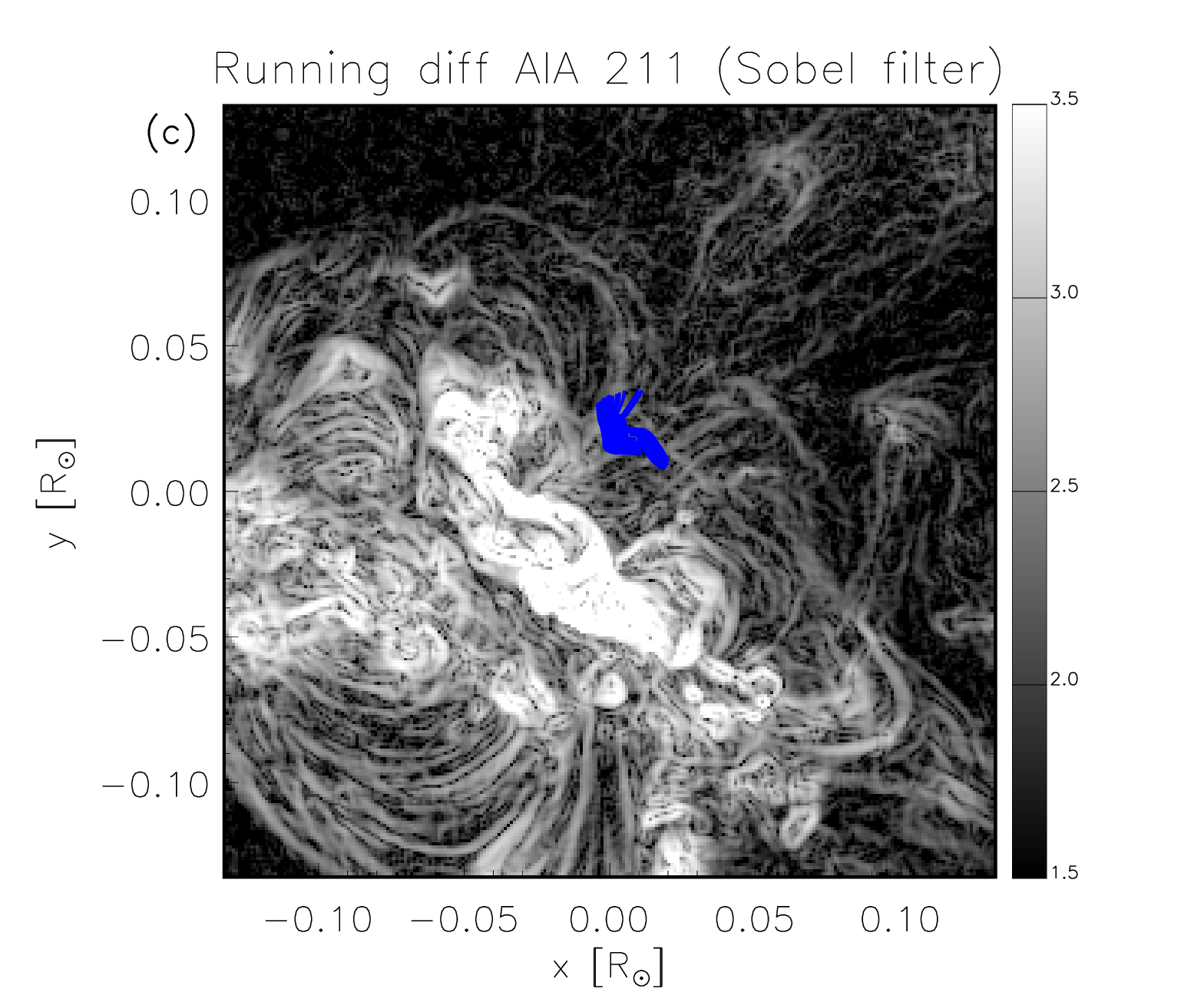}
}

\caption{Maps of the surface magnetic field measured on (a) 31 July 2011 at 05:00:41 UT
(the heliographic coordinates of the center of the image are $x$~= -252$^{\prime\prime}$, $y$~= 178$^{\prime\prime}$)
and (b) on 2 August 2011 at 06:00:41 UT (the center of the image is at $x$~= 174$^{\prime\prime}$, $y$~= 188$^{\prime\prime}$).
In (b) we overplot some magnetic field lines from the 3D magnetic configuration,
obtained with the NLFF time series. (c) Map of running difference images from AIA 211~{\AA} filter with a Sobel filter applied
(scale of difference in DN) on 2 August 2011 at 06:00:41 UT. Blue lines are representative of the flux rope.
}
\label{nlfffevol}
\end{figure}

In Figure \ref{nlfffevol}a--b we show the initial magnetograms on 31 July 2011 at 05:00:41 UT
and the final magnetogram on 2 August 2011 at 06:00:41 UT with the
final 3D magnetic configuration, shown by the green lines.
The initial magnetic configuration is assumed to be potential,
while the final stage is highly non-potential with a
magnetic flux rope formed.
Over the time series of NLFF fields a magnetic flux rope forms
as a consequence of the continuous motion and evolution of the magnetic field at the lower boundary.
The flux rope forms along the polarity inversion line (PIL) and is about 0.03~R$_{\odot}$
long; its central point is located at the coordinates $x$ = 0.0078~R$_{\odot}$ and $y$ = 0.018~R$_{\odot}$, which in the MHD
simulation domain correspond to the heliographic coordinates
$x$~= 181$^{\prime\prime}$, $y$~= 205$^{\prime\prime}$ on 2 August 2011 at 06:00:41 UT.
The flux rope covers only part of the PIL,
whereas sheared magnetic field lines are present along the whole PIL,
marking the non potentiality of the final magnetic field configuration.

Figure \ref{nlfffevol}c shows a running difference of AIA 211~{\AA} pass band image
at the same time of the last magnetogram where we apply the sobel filter
to highlight the coronal structures probably associated with the expanding flux rope.
We find that the general topology of the loops is reproduced
where the magnetic field lines match
the loop structures at the left-lower corner of the image,
the system of loops depart in different directions from the enhanced emission region,
and many of the loops around the magnetic flux rope.

It is also crucial to point out that due to a slow variation of the magnetic field,
the final magnetic configuration is not sensitive to time intervals
of some minutes in comparison to the 60 minutes sampling time of the HMI magnetograms.
This means that the final configuration could represent any
time within some minutes around the exact magnetogram used.

\subsection{MHD Model} 
The final 3D magnetic configuration from the NLFF time series is then used
as input for an ideal MHD simulation.
This approach is an extension of the technique successfully adopted in
\citet{pagano13b} and then further developed in \citet{pagano13a}
and \citet{pagano14}, where the magnetic configuration, obtained
from a magneto-frictional non-linear force-free model, is input as initial condition in an ideal MHD simulation.

In order to do so,
we import the three components of the magnetic field from the NLFF time series grid
to the grid of our 3D MHD model.
Specifically in \citet{pagano13b} and \citet{pagano13a} there are a number of thorough tests
to show that this is possible so that we preserve
the stability or instability of the configuration.

\subsubsection{Interpolation of the Magnetic Configuration} 
  \label{S-simple-equations}
In the present article we have simplified the way in which the 3D interpolation is performed and
have adopted a cartesian grid.
In cartesian coordinates, let $B(x,y,z)$ be the value that we want to compute in the position $(x,y,z)$
that we know lies in the cell, defined by the indexes $[i:i+1,j:j+1,k:k+1]$, where the quantity $b$ is defined.
We compute
\begin{equation}
B(x,y,z)=\sum_{i,j,k}^{i+1,j+1,k+1} b(i,j,k) V(i,j,k) / V
\end{equation}

where $V(i,j,k)$ is the volume, defined by the point $(x,y,z)$ and the cell corner opposite to the position $(i,j,k)$, and
V is the sum of these volumes.
This approach guarantees the continuity of the solution and its smoothness independently of the spatial resolution of the
grid where $B(x,y,z)$ is defined.

\subsubsection{Plasma Distribution}
\label{plasmadistribution}
As the time series of NLFF fields provides only the magnetic configuration,
we need to adopt
an initial distribution of plasma density, speed, and temperature
in order to close a complete set of MHD variables.
We aim at a realistic and general representation of the solar corona, so
we seek to produce a distribution of plasma that takes into account the heterogeneity of the coronal plasma.
In particular, we intend to describe the initial flux rope plasma as colder and denser than the plasma outside of the flux rope.
Magnetic flux ropes are structures where the magnetic field is usually
more intense than in their surroundings and where the magnetic field presents twist.
Therefore, we define the following proxy function to link the plasma temperature and density
to the magnetic field:

\begin{equation}
\label{omegab}
\omega=\sqrt{\omega_x^2+\omega_y^2+\omega_z^2}
\end{equation}
where
\begin{equation}
\label{omegax}
\omega_x=\frac{\left|B\times\nabla B_x\right|}{\left|\nabla B_x\right|}
\end{equation}
\begin{equation}
\label{omegay}
\omega_y=\frac{\left|B\times\nabla B_y\right|}{\left|\nabla B_y\right|}
\end{equation}
\begin{equation}
\label{omegaz}
\omega_z=\frac{\left|B\times\nabla B_z\right|}{\left|\nabla B_z\right|}
\end{equation}

where $\vec{B}=(B_x,B_y,B_z)$ is the magnetic field, the cartesian coordinates $x$ and $y$ are tangent to the solar surface and
$z$ is perpendicular to the surface.
The function $\omega$ is positive definite and peaks where the magnetic field presents more twist and is more intense,
\textit{e.g.} near the centre of the magnetic flux rope axis.
As an example, at the flux rope axis $\nabla B_z$ is parallel to the solar surface
along the direction connecting two opposite polarities
and perpendicular to the magnetic field that is mostly parallel to the $x$\,--\,$y$ plane.
In this configuration $\omega_z$ is relatively high.
Additionally, the value of $\omega$ is proportional to the magnetic field intensity,
which results in it being higher near the solar surface
and lower at further radial distances from the solar surface.

In order to effectively use $\omega$ to model the solar atmosphere we define the function:
\begin{equation}
\Omega=\arctan{\frac{\omega-\omega^{\star}}{\Delta\omega}}/\pi+0.5
\end{equation}
where $\omega^{\star}$ and $\Delta\omega$ are two simulation parameters.
$\Omega$ is then bound between $0$ and $1$ and the temperature is defined by:
\begin{equation}
\label{tempomega}
T=\Omega(T_{flux rope}-T_{corona})+T_{corona}
\end{equation}
where $T_{\mathrm{flux rope}}$ and $T_{\mathrm{corona}}$ are two simulation parameters
that respectively represent the temperature in the flux rope and in the external corona.
The thermal pressure is independently specified by the exponential solution
for hydrostatic equilibrium with constant gravity
with a uniform temperature set equal to $T_{\mathrm{corona}}$:
\begin{equation}
\label{presgrav}
p=\frac{\rho_0}{\mu m_p}k_B T_{\mathrm{corona}} \exp{\left(-z\frac{g \mu m_p}{k_B T_{\mathrm{corona}}}\right)}
\end{equation}
where $p$ is the thermal pressure, $\rho_0$ is a simulation parameter that
sets the density at the solar surface, $\mu=$1.31 is the average particle mass, $m_p$ is the proton mass,
$k_B$ is the Boltzmann constant and $g$ is the solar gravitational acceleratation at the solar surface.
The density $\rho$ is given by the equation of state:
\begin{equation}
\label{eos}
\rho=\frac{p}{T(\vec{B})}\frac{\mu m_p}{k_B}
\end{equation}
where $\rho$ is the density,
$T$ is the temperature.

\subsubsection{MHD Simulation} 
  \label{S-long-equations}
Based on the approach described in Section \ref{plasmadistribution},
and using the final 3D magnetic field configuration,
obtained from the NLFF time series as described in Section \ref{S-equations},
we construct the initial condition for the MHD simulation.
Table \ref{tableparameters} shows the values used in our model for the relevant parameters.

\begin{table}   
\caption{Parameters for the initial condition of the ideal MHD simulation}             
\label{tableparameters}      
\begin{tabular}{c c c}        
\hline
Parameter & Value & Units  \\    
\hline                        
   $\rho_0$ & $1.1\times10^{-12}$ & g cm$^{-3}$  \\      
   $\omega^{\star}$ & $300$  & $G$ \\
   $\Delta\omega$ & $80$  & $G$ \\
   $T_{\mathrm{flux rope}}$ & $10^5$ & K \\
   $T_{\mathrm{corona}}$ & $2\times10^6$ & K \\

\hline                                   
\end{tabular}
\end{table}

We use the MPI-parallelized Adaptive-Mesh Refinement Versatile-Advection Code
(MPI-AMRVAC) software~\citep{porth14}, to solve the MHD equations,
where external gravity is included as a source term,
\begin{equation}
\label{mass}
\frac{\partial\rho}{\partial t}+\vec{\nabla}\cdot(\rho\vec{v})=0,
\end{equation}
\begin{equation}
\label{momentum}
\frac{\partial\rho\vec{v}}{\partial t}+\vec{\nabla}\cdot(\rho\vec{v}\vec{v})
   +\nabla p-\frac{(\vec{\nabla}\times\vec{B})\times\vec{B})}{4\pi}=\rho\vec{g},
\end{equation}
\begin{equation}
\label{induction}
\frac{\partial\vec{B}}{\partial t}-\vec{\nabla}\times(\vec{v}\times\vec{B})=0,
\end{equation}
\begin{equation}
\label{energy}
\frac{\partial e}{\partial t}+\vec{\nabla}\cdot[(e+p)\vec{v}]=\rho\vec{g}\cdot\vec{v},
\end{equation}
where $t$ is time, $\vec{v}$ velocity, $\vec{g}$ is the vector of the solar gravitational acceleration,
and the total energy density, $e$, is given by
\begin{equation}
\label{enercouple}
e=\frac{p}{\gamma-1}+\frac{1}{2}\rho\vec{v}^2+\frac{\vec{B}^2}{8\pi}
,\end{equation}
where $\gamma=5/3$ denotes the ratio of specific heats.

The computational domain is composed of $256\times256\times248$ cells, distributed on a uniform grid.
The simulation domain is similar to the one used in the time
series of NLFF fields and it extends over 0.267~R$_{\odot}$ in the $x$ and $y$ direction
and over 0.266~R$_{\odot}$ in the $z$ direction, starting from $z$~= 0.0027~R$_{\odot}$ above
the photosphere. The boundary conditions are treated with a system of ghost
cells. Two layers of cells between $z$~= 0 and $z$~= 0.0021~R$_{\odot}$ are used as fixed
lower boundary conditions during the ideal MHD simulation.
Open boundary conditions are imposed at the outer boundary
and finally reflective boundary conditions are set at the $x$ and $y$ boundaries of the simulation box.

\begin{figure}   
\centerline{\hspace*{0.005\textwidth}
\hspace*{-0.02\textwidth}
\includegraphics[scale=0.2,viewport=0 40 700 550,clip=]{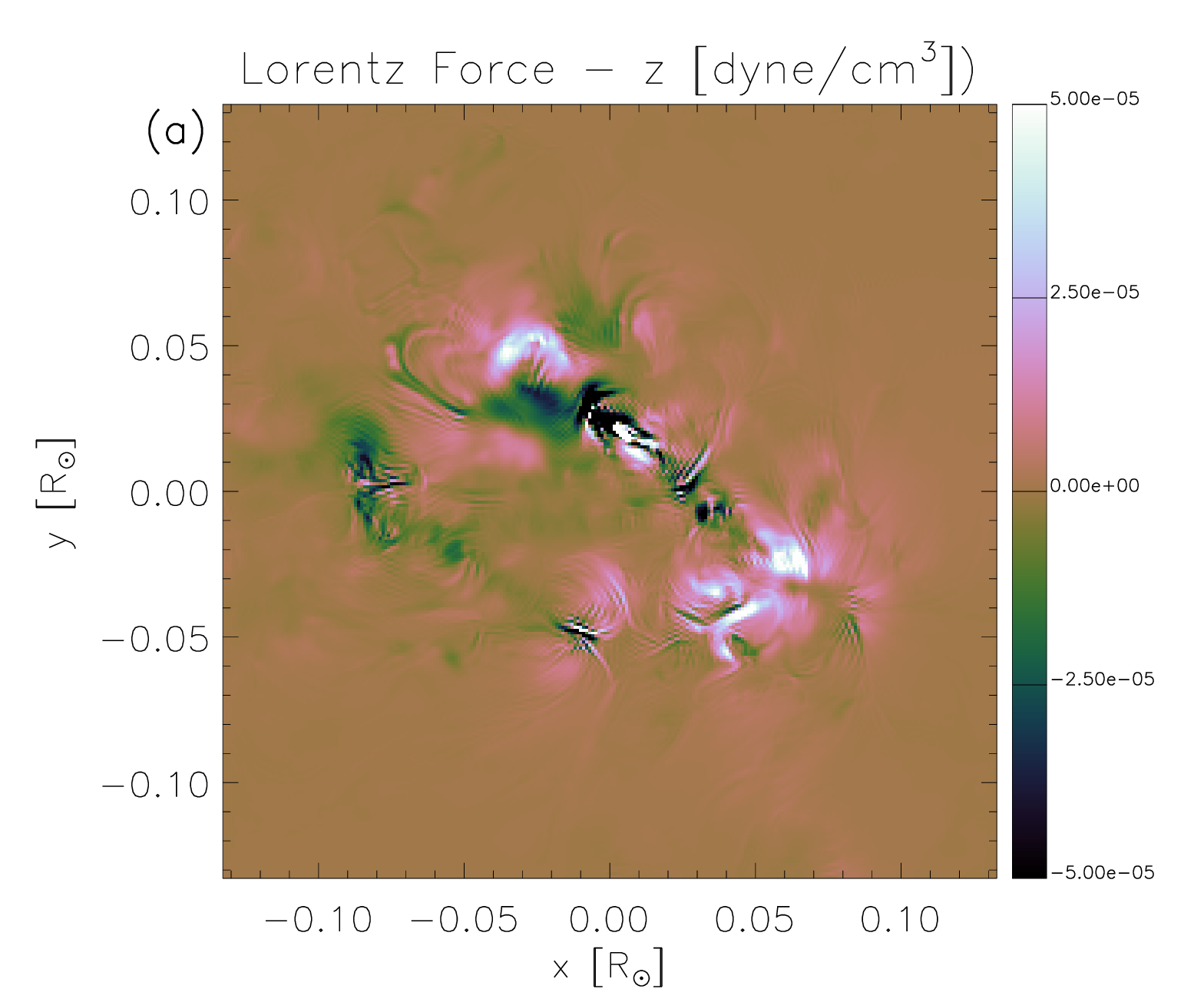}
\hspace*{-0.02\textwidth}
\includegraphics[scale=0.2,viewport=55 40 700 550,clip=]{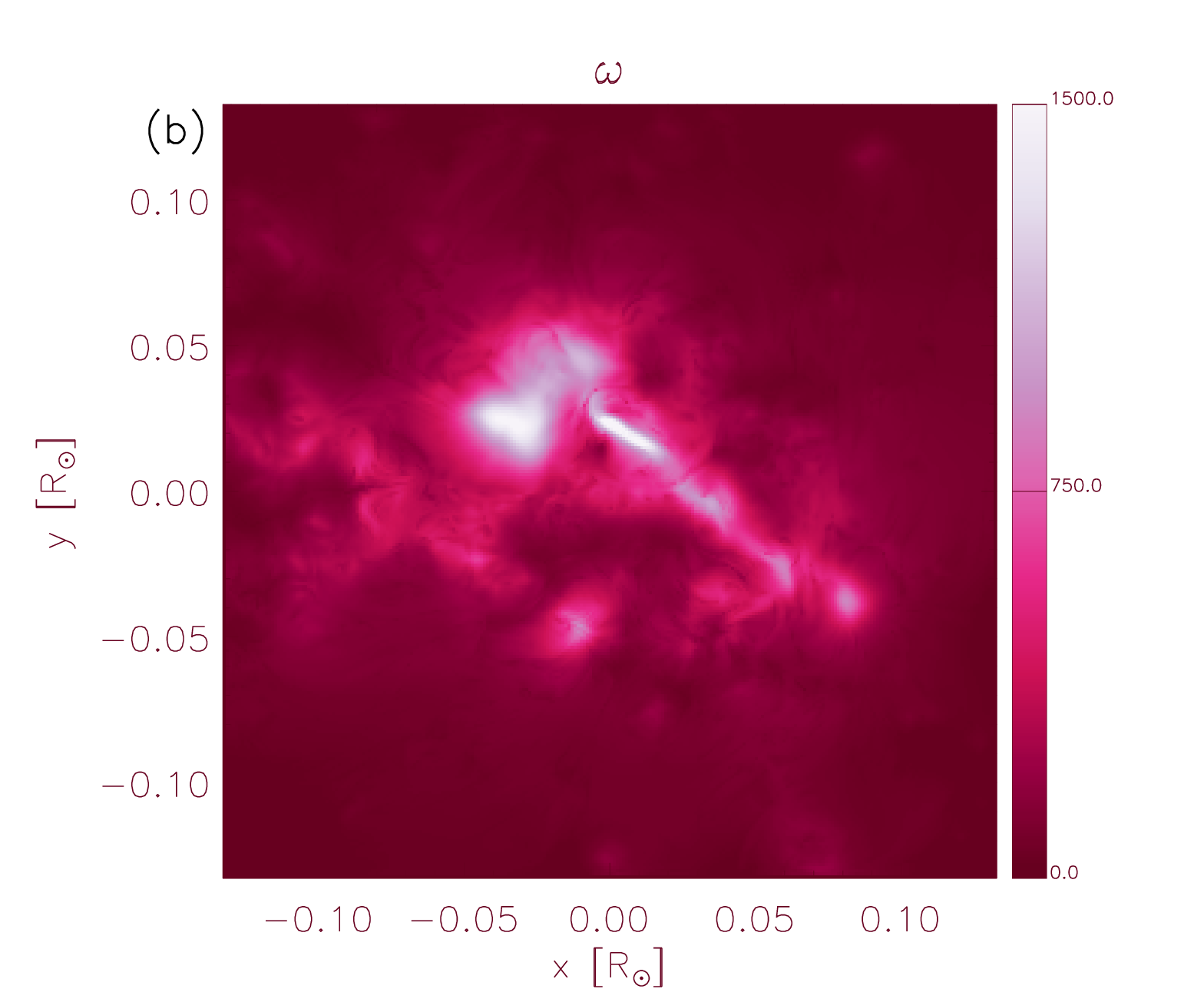}
\vspace*{0.01\textwidth}}
\centerline{\hspace*{0.005\textwidth}
\hspace*{-0.02\textwidth}
\includegraphics[scale=0.2,viewport=0 0 700 550,clip=]{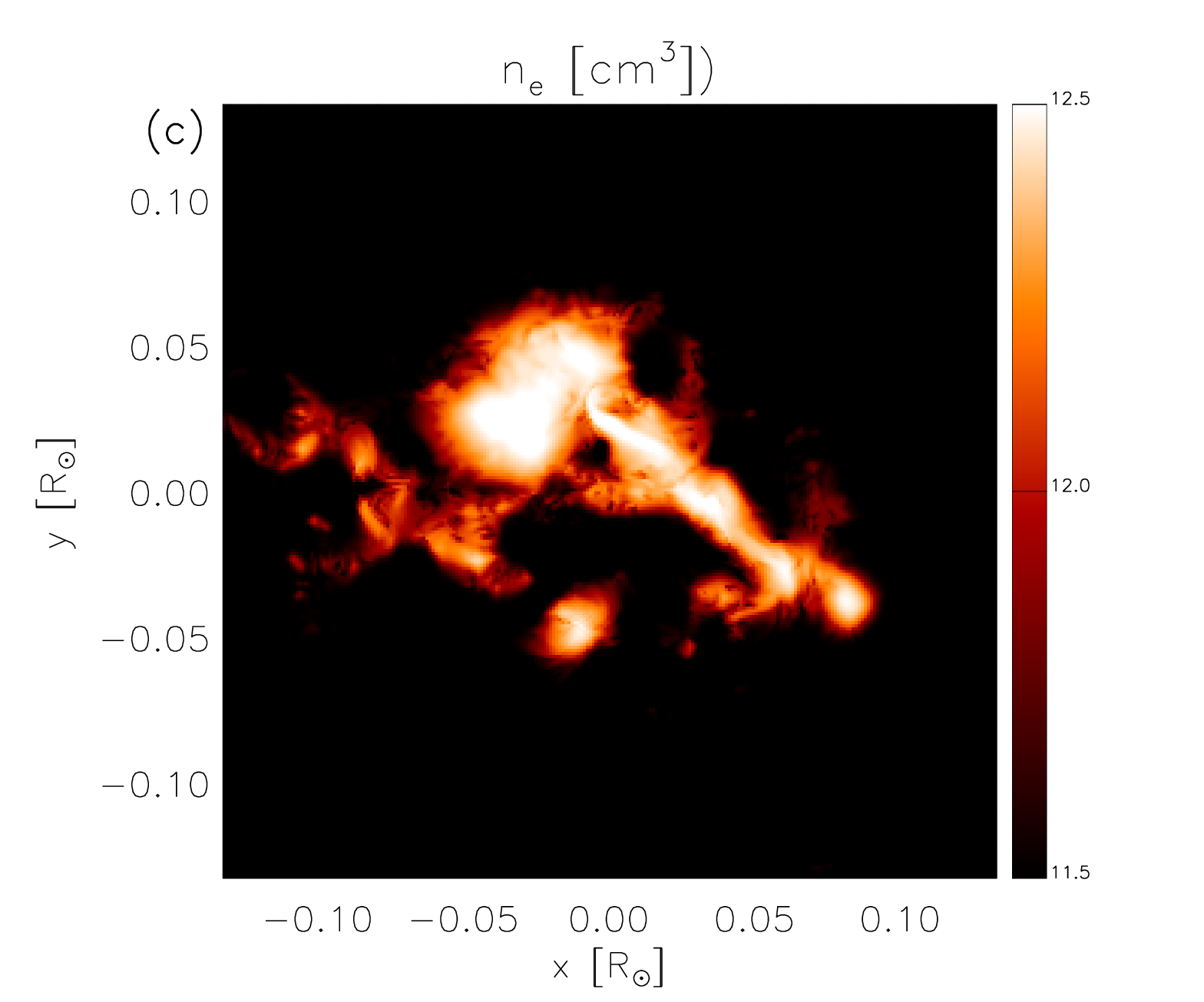}
\hspace*{-0.02\textwidth}
\includegraphics[scale=0.2,viewport=55 0 700 550,clip=]{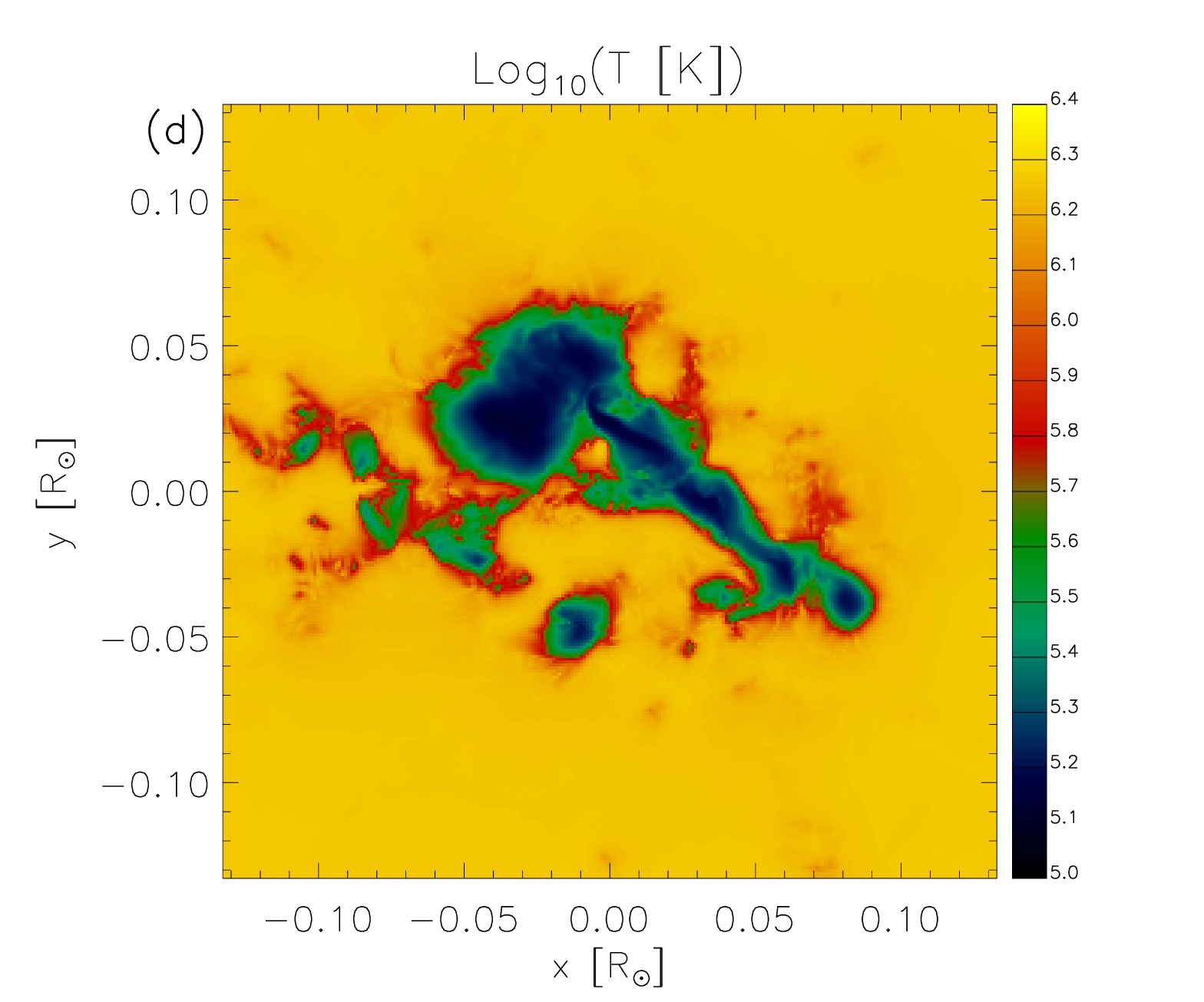}
}
\caption{Lower boundary of the initial condition of the ideal MHD simulation:
 (a) Lorentz Force,  (b) $\omega$, (c) Log$_{10}$($\rho$ [g cm$^{-3}$]), and (d) Log$_{10}$ (T [K]). The axes scales are in units of R$_{\mathrm{sun}}$.
 }
\label{initialcond}
\end{figure}

Figure \ref{initialcond} shows the values of the radial component of the Lorentz force at the lower boundary,
the function $\omega$, derived from the magnetic configuration,
and the resulting distribution of electron density $n_e$ and temperature $T$.
The Lorentz force is maximum at the location of the flux rope.
The map of $\omega$ follows the pattern we have prescribed,
being higher around the region of the PIL and peaking where the flux rope is located.
Consequently, the maps of $n_e$ and T show the regions with a higher twisted magnetic field.
In particular, the position where the magnetic flux rope sits
(compare with Figure \ref{nlfffevol})
presents a density $\sim10$ times higher than its surroundings and
a temperature value $\sim10$ times lower.
It is also to be noted that in our model the location, where the flux rope is relatively cold,
is consistent with the observations of AIA in the 211~{\AA} band (Figure \ref{nlfffevol}c)
as the location where the flux rope is not visible, since the filter is tuned to observe plasma at higher temperatures.

The present initial conditions are clearly out of equilibrium for a number of reasons.
In our simulation the initial plasma $\beta$ ranges between the two extrema of
$\beta\sim10^{-3}$ (at the flux ropes)
and $\beta\sim10^3$ (in very confined regions where the magnetic field is less intense).
Elsewhere it lies between 10$^{-2}$ and 10$^{-1}$.
Therefore, the strongest unbalanced force in the initial condition is the Lorentz force in the magnetic flux rope, which is upwards.
At the same time, the radial profiles of density and pressure
do not prescribe the balance between the thermal pressure gradient and the gravity force.
However, as addressed in detail in \citet{pagano13b}, such unbalance shows effects over timescales longer
than the dynamics triggered by the Lorentz force and therefore these can be neglected.

\section{Results of the MHD Simulation and Comparison with Observations}
\label{simulation}
We run the MHD simulation starting from the final magnetic configuration,
obtained from the time series of NLFF fields, when the magnetic flux rope is fully formed,
in order to describe the evolution of the ejection of the magnetic flux rope.
The pre-eruptive initial magnetic configuration described in Section~4.1
is obtained from a long term sequence of magnetograms with 1 hour cadence
and thus it cannot indicate the exact magnetic flux rope ejection start time.
Therefore we present the MHD evolution in terms of the
time elapsed from the initial condition and
we impose the time of the MHD simulation initial condition
to match the observed CME initiation time.
For this specific simulation, in Table \ref{T-Flares} we report
the onset of the eruption, \textit{i.e.} the time of the intial conidition, at 05:54:40 UT.
The initial condition of the MHD simulation is out of equilibrium and several
plasma flows occur at the beginning.

\begin{figure}  
\centerline{\hspace*{0.05\textwidth}
\includegraphics[scale=0.35,viewport=0 40 360 250,clip=]{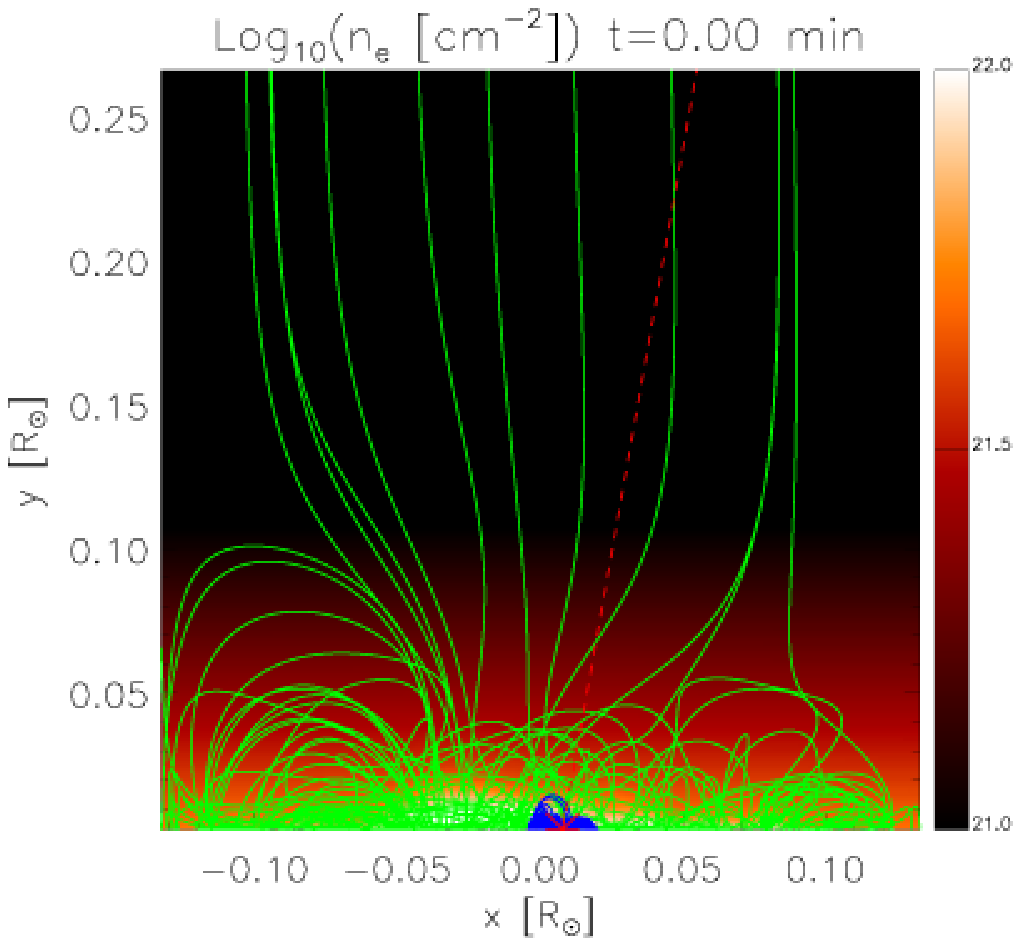}
\hspace*{-0.02\textwidth}
\includegraphics[scale=0.35,viewport=65 40 360 250,clip=]{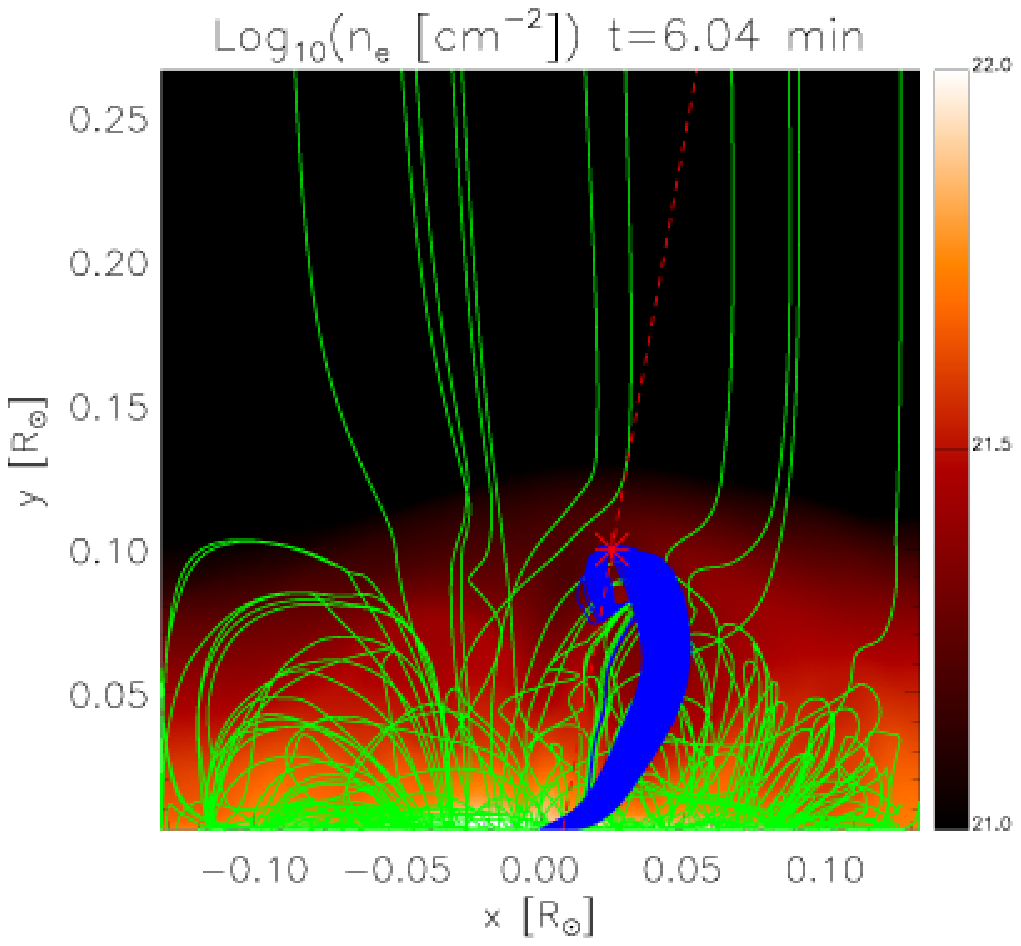}
\hspace*{-0.02\textwidth}
\includegraphics[scale=0.35,viewport=65 40 360 250,clip=]{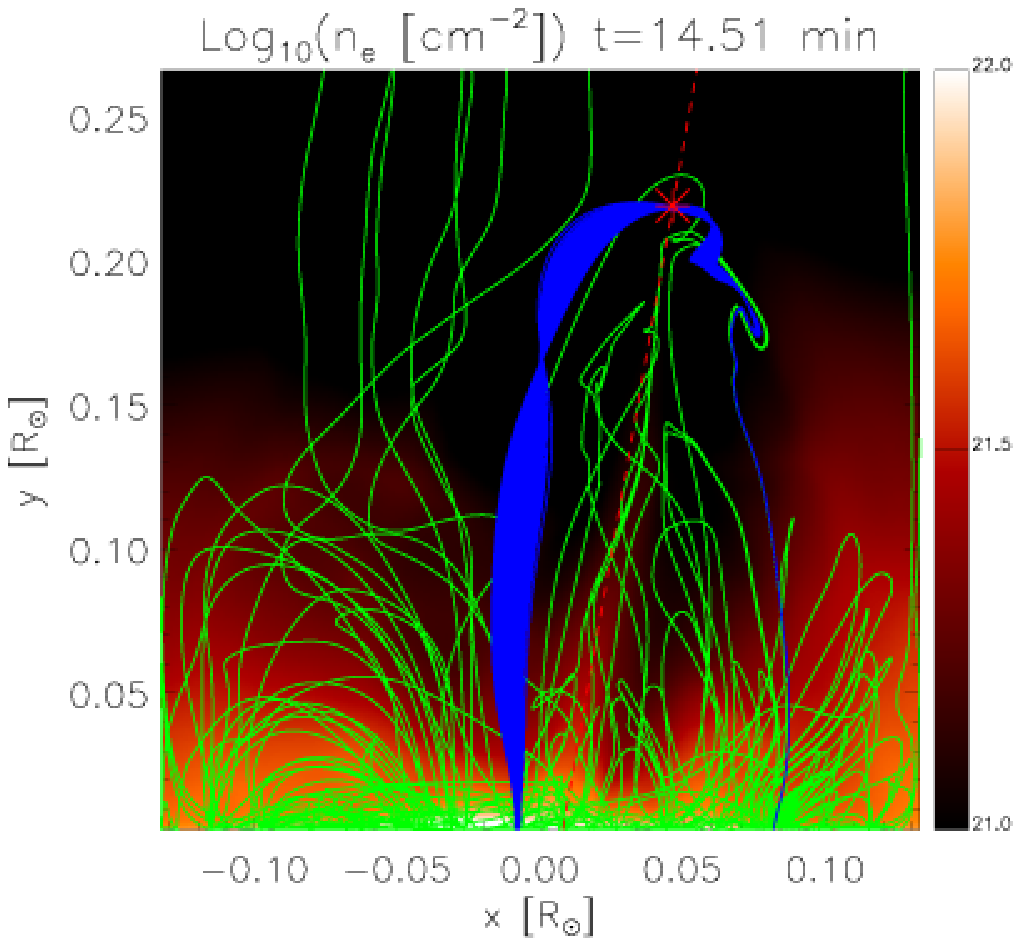}
\vspace*{0.01\textwidth}
}
\centerline{\hspace*{0.05\textwidth}
\includegraphics[scale=0.17,viewport=0 40 700 550,clip=]{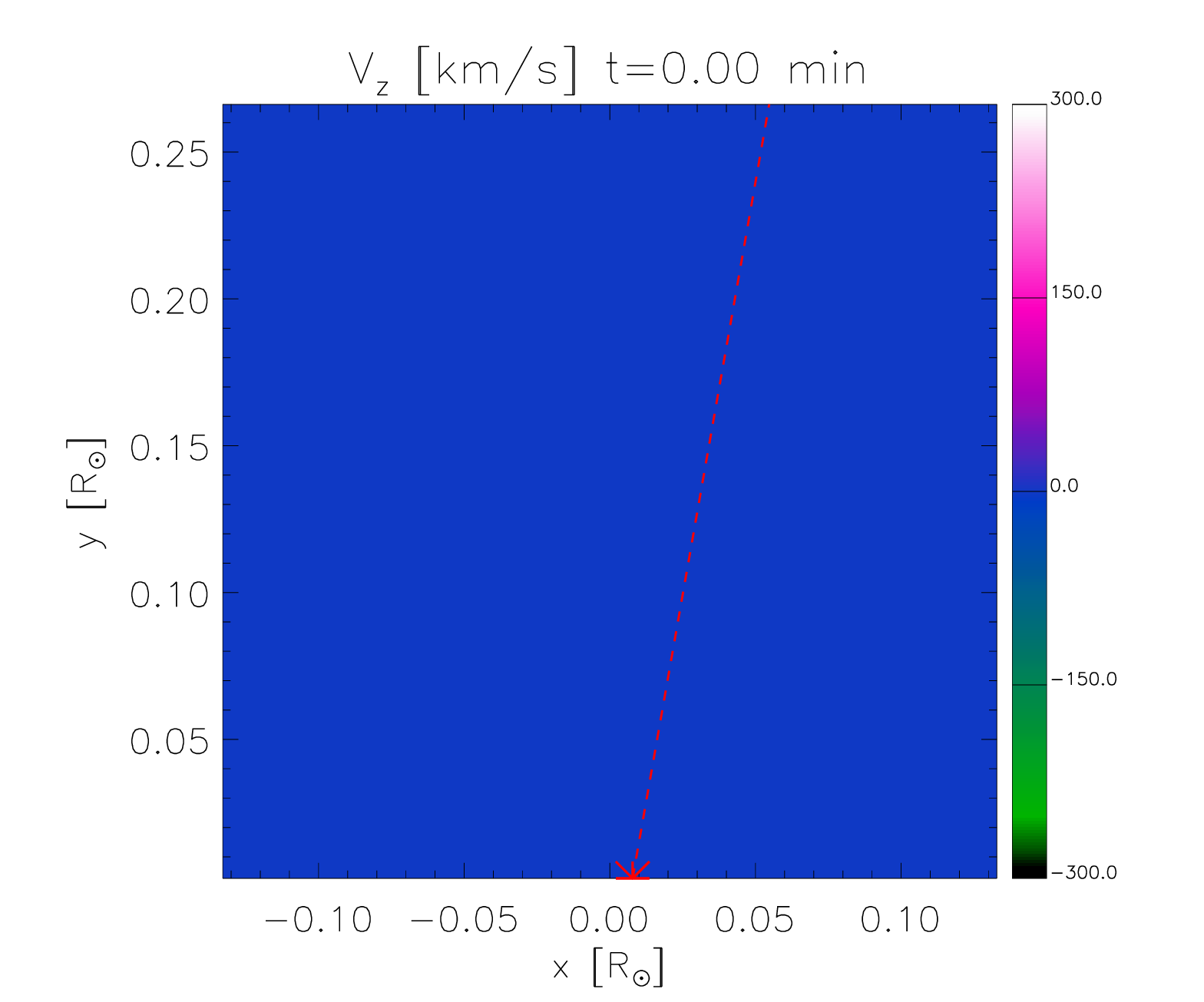}
\hspace*{-0.02\textwidth}
\includegraphics[scale=0.17,viewport=65 40 700 550,clip=]{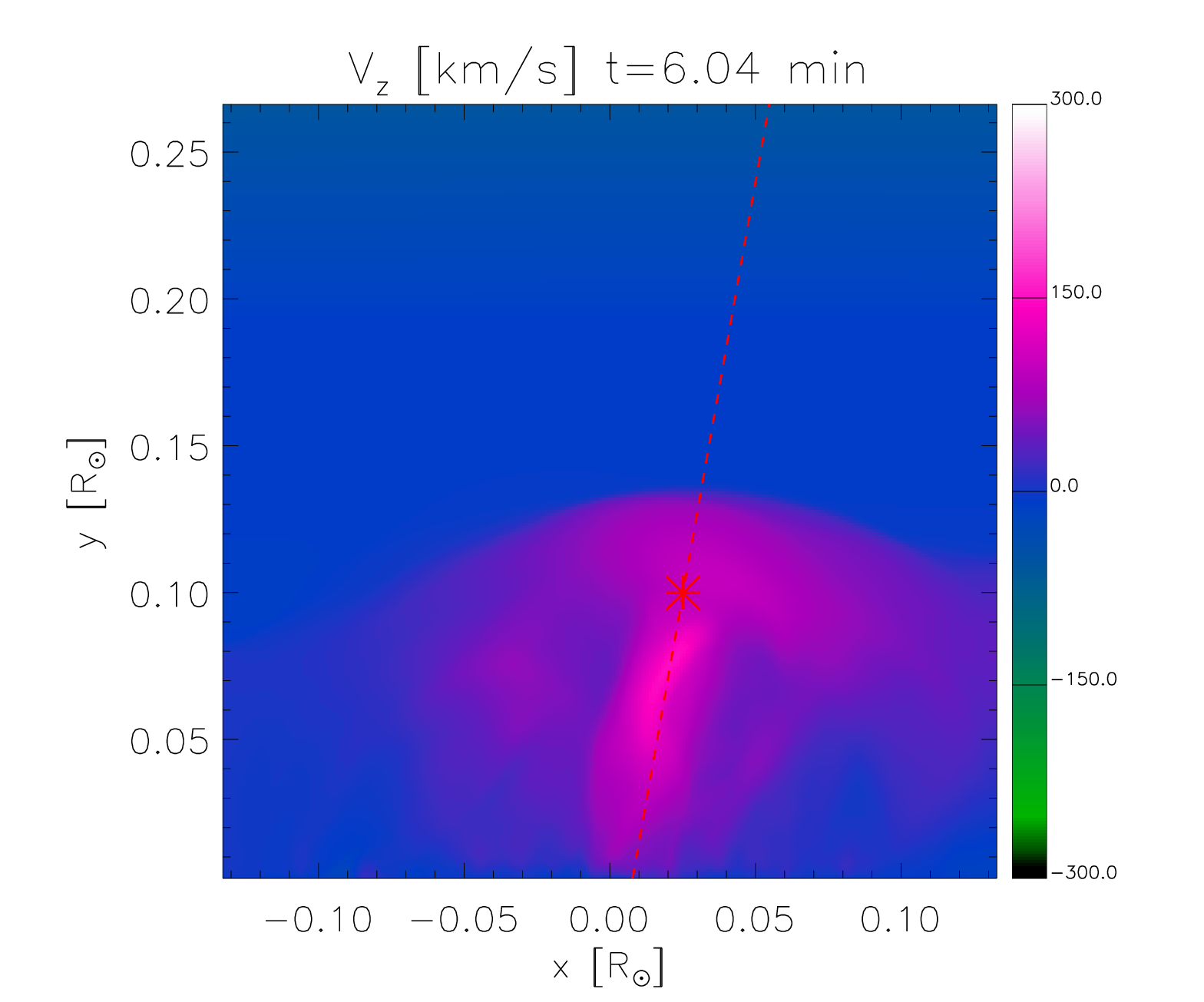}
\hspace*{-0.02\textwidth}
\includegraphics[scale=0.17,viewport=65 40 700 550,clip=]{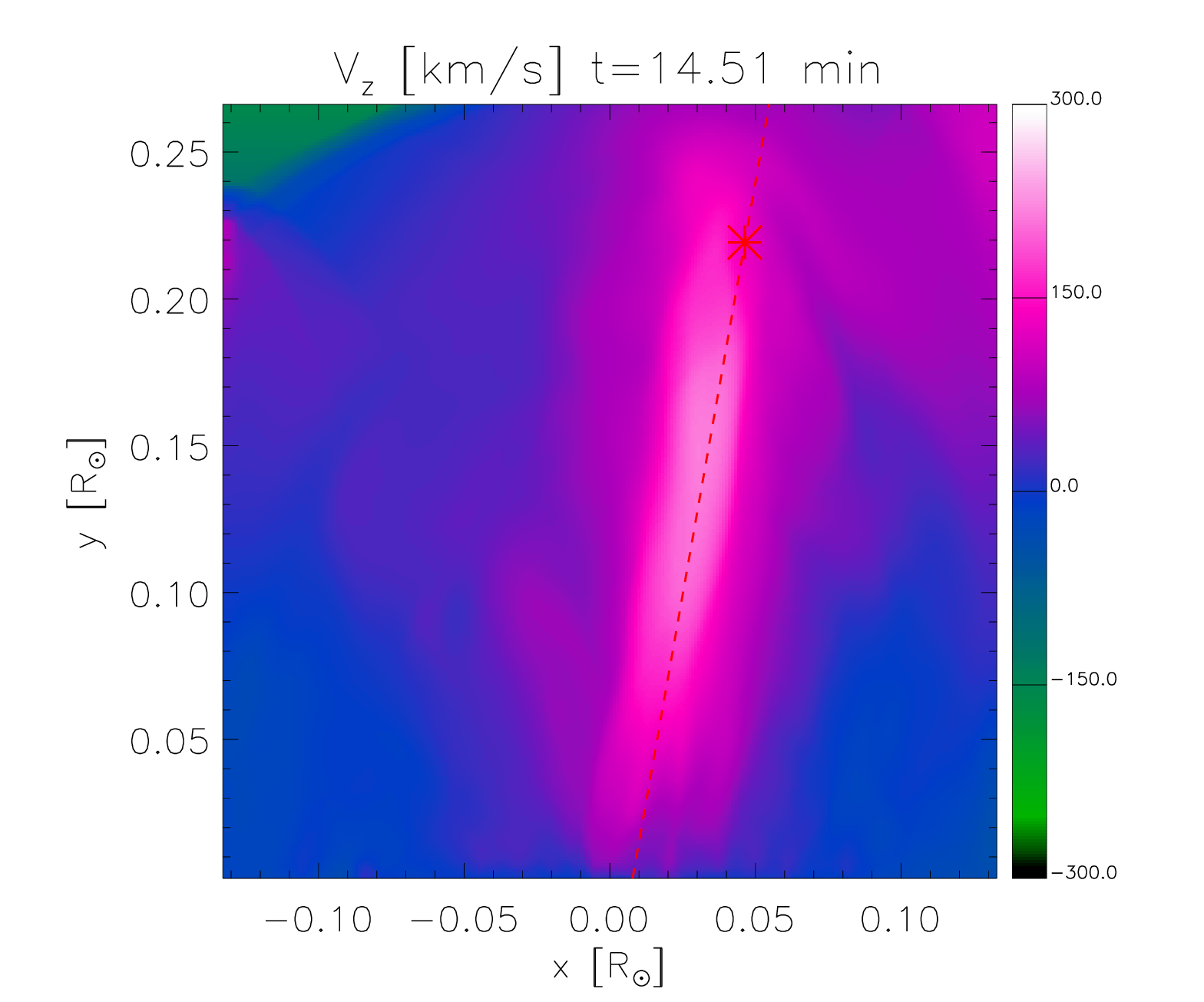}
}
\vspace*{0.01\textwidth}
\centerline{\hspace*{0.05\textwidth}
\includegraphics[scale=0.17,viewport=0 0 700 550,clip=]{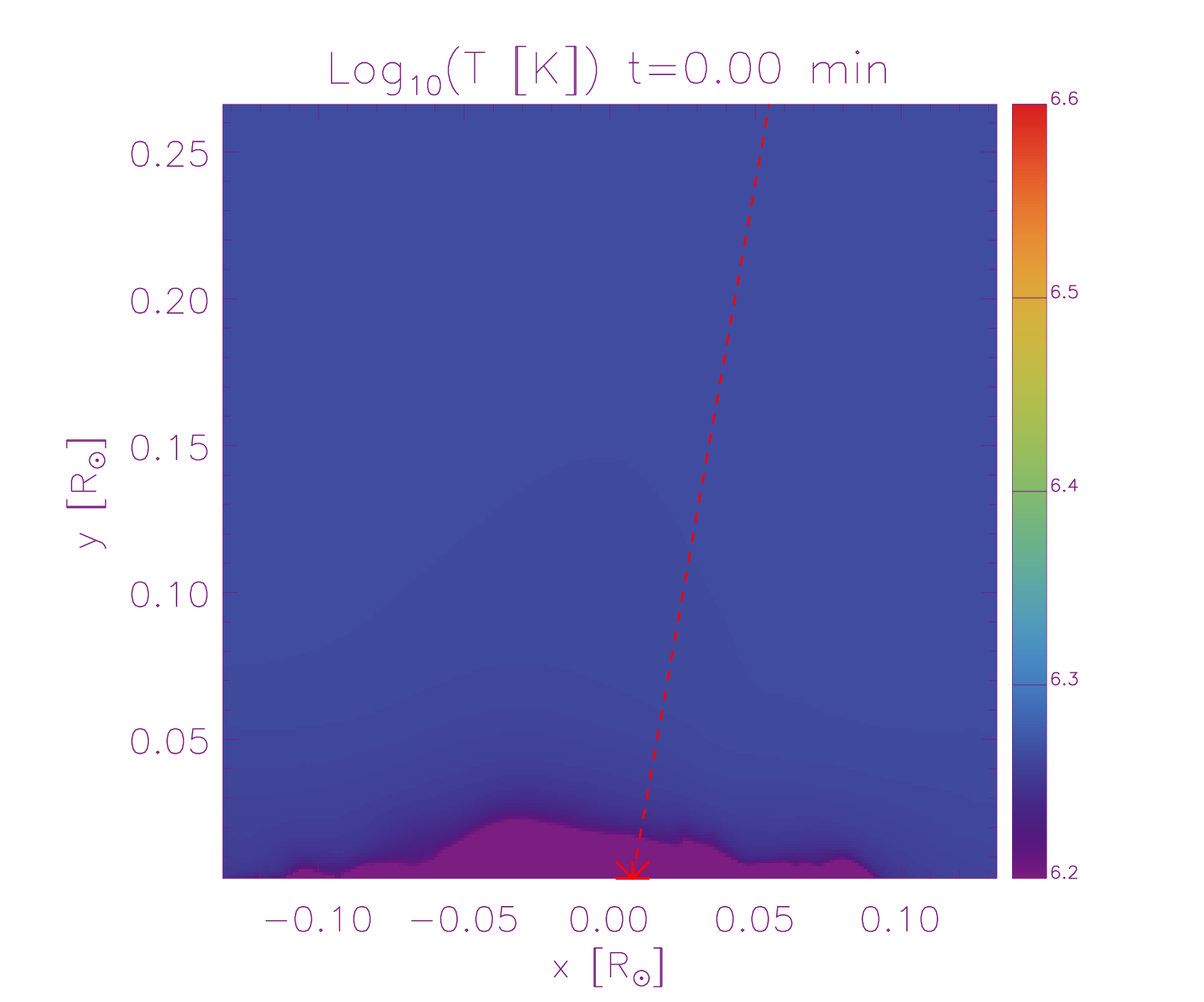}
\hspace*{-0.02\textwidth}
\includegraphics[scale=0.17,viewport=65 0 700 550,clip=]{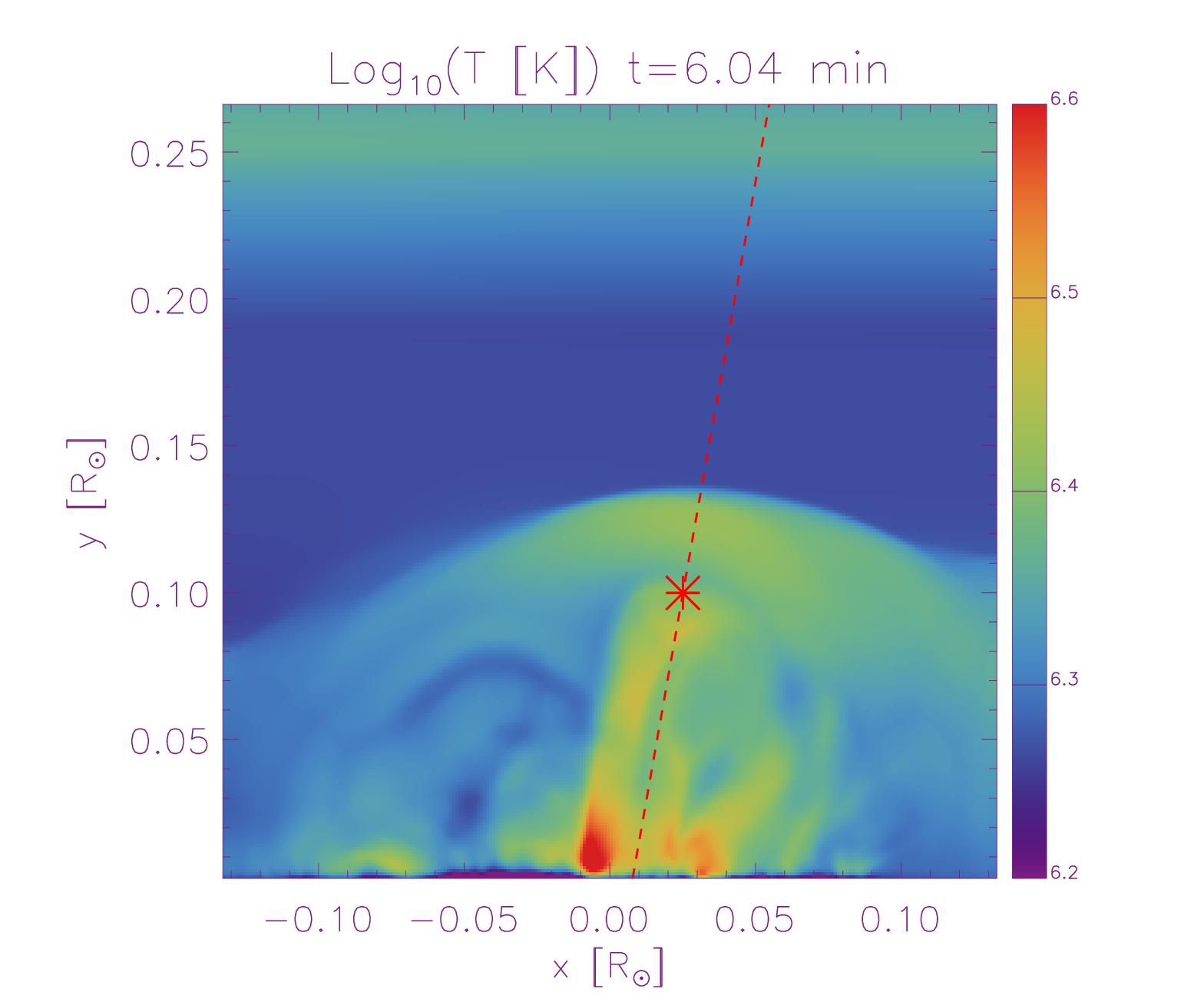}
\hspace*{-0.02\textwidth}
\includegraphics[scale=0.17,viewport=65 0 700 550,clip=]{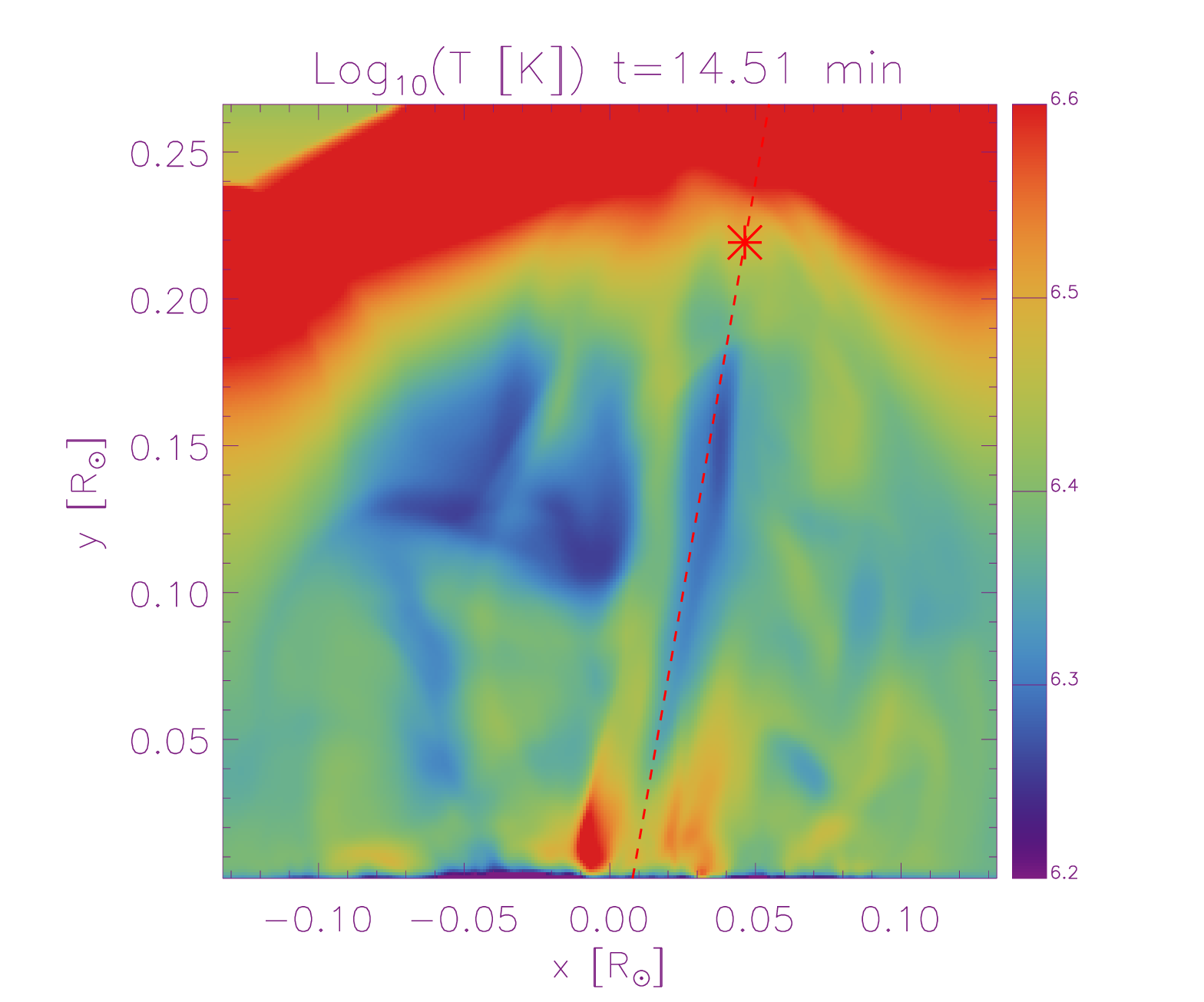}
}

\caption{Panels showing the evolution of the MHD simulation. The upper row shows the integral of the electron density,
the middle row, and the lower row show, respectively, the integrals of the $z$ component of the
velocity and temperature averaged by the electron density.
In the upper row some magnetic field lines (green lines) and flux rope field lines (blue lines) are overplotted.
The left column shows quantities at the time in the simulation corresponding to
$t$~= 0 min, the central column to $t$~= 6.04 min, and the right column to $t$~= 14.51 min from the eruption onset.
The red dashed line represents the direction of propagation of the magnetic flux rope
and the red star is the position of the centre of the flux rope at each time.}
\label{evol}
\end{figure}

However, the dominant evolution occurs where the flux rope plasma is pushed upwards
by the unbalanced Lorentz force and the flux rope starts erupting.
In order to follow the evolution of the flux rope, we display the simulation
from a line of sight parallel to the $y$ axis.
In Figure \ref{evol} we show the electron density, integrated along this line of sight,
together with the temperature and velocity in the $z$ direction, integrated along the line of sight
and weighted over $n_e$.
We use as characteristic stages in the evolution the times
of $t$~= 0 min, $t$~= 6.04 min, and $t$~= 14.51 min from the eruption onset.
Additionally, by considering the evolution of $v_z$ and $\omega$, we manually track
the directions of propagation of the magnetic flux rope and the position of its centre along this direction
at each snapshot of the simulation.
This is possible, because the cuts of both $v_z$ and $\omega$
along the direction of propagation of the flux rope show a local peak at the flux rope centre.
This direction is represented by the red line and the position of the red star in Figure \ref{evol}.

The particle density maps show an expansion and an upwards propagation of the structures
that are initially lying low in the domain.
The same motion is highlighted by the visible change in the magnetic field lines.
In particular, the flux rope magnetic field lines (represented by the blue lines) show a distinctive motion,
where the magnetic field lines are increasingly longer and less twisted over the evolution.

The evolution of $v_z$ describes well this behaviour,
where we see a region of the upward directed velocity that is
composed of an elongated region where the plasma speed is higher and a bow-shaped region ahead
that represents the front of the ejecta.
In particular, the elongated region extends along the direction of propagation of the flux rope
and dominates over other motions present in the domain.
The maximum of $v_z$ reaches values greater than 500~km~s$^{-1}$ at $t$~= 11.5 min in the MHD simulation.
It should be noted that some boundary effects are visible at the external boundary,
where some moderate inflow develops over the course of the simulation.
However, the low density of the inflowing plasma makes this effect negligible for the dynamics of the ejection.
It is interesting to notice that no shocks are formed ahead of the flux rope ejection.
If we take into consideration the MHD simulation snapshot at 06:06:43 UT,
we find that the plasma speed at the front of the ejection is $\sim$~130~km~s$^{-1}$,
where the sound speed is $\sim$~300~km~s$^{-1}$.
Only behind the propagation of the flux rope,
we find regions with supersonic or superalfvenic velocities,
where the Mach number is 1.5 or the Alfv\'enic Mach number is $\sim$~3.

The evolution of the temperature is significantly complex due to the energetics of the flux rope ejection.
As already shown in similar simulations \citep{pagano14}, the numerical resistivity plays a crucial role in heating the plasma,
while expansion and decompression can lead to the cooling of the plasma in certain regions.
As soon as the simulation starts, magnetic energy is converted into thermal energy and this leads
to an overall increase of the temperature.
At the same time, this is not happening everywhere, but only where the magnetic field is initially more twisted, \textit{e.g.} near the PIL.
Subsequently, the temperature decreases in some regions, but overall stays above the initial one.

\begin{figure}  
\centerline{\hspace*{0.005\textwidth}
\includegraphics[scale=0.5,viewport=0 0 350 250,clip=]{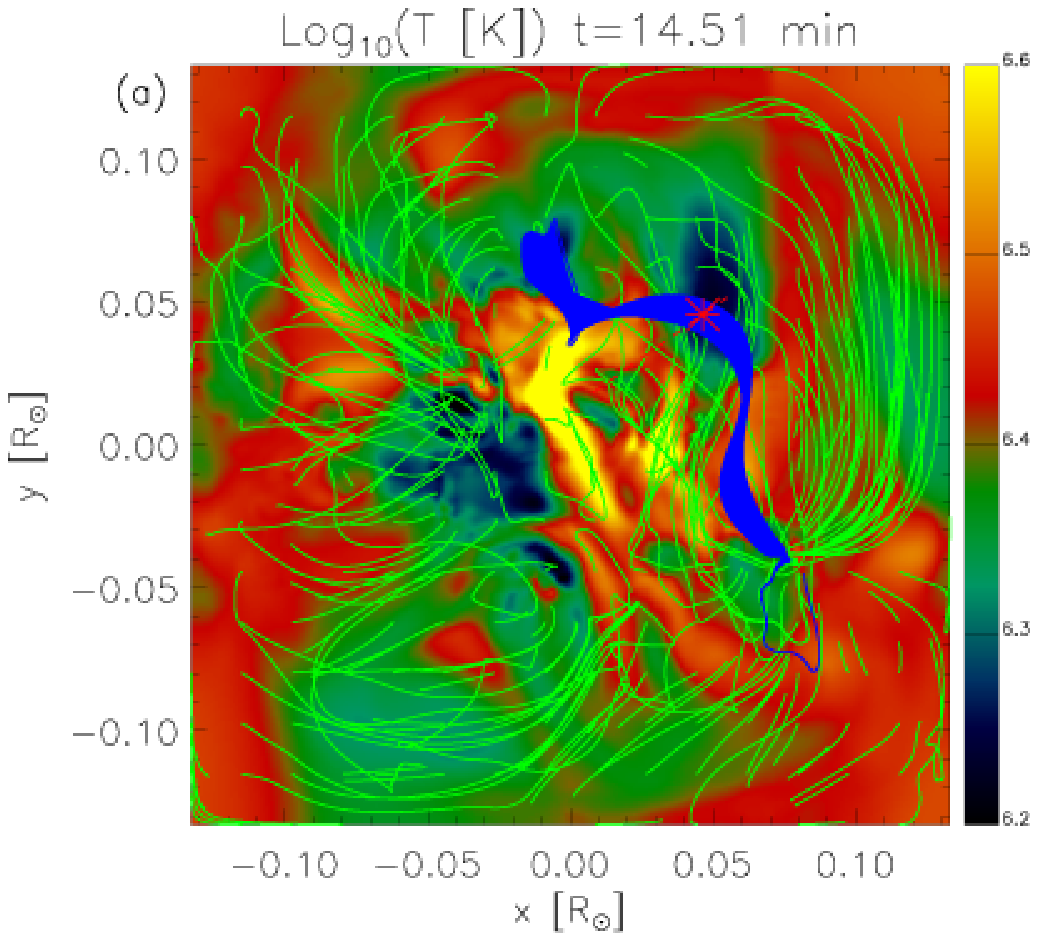}
\includegraphics[scale=0.25,viewport=55 0 700 550,clip=]{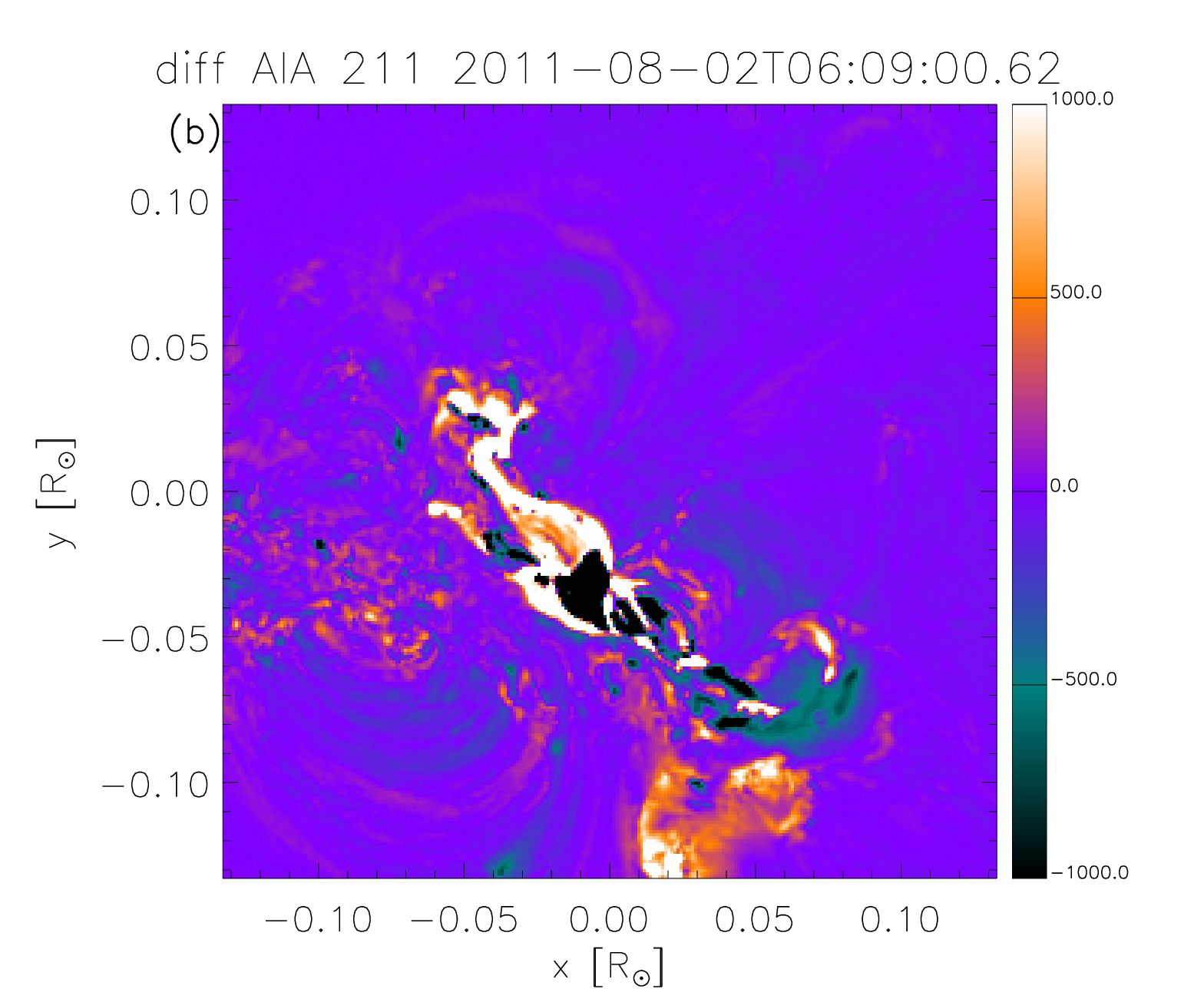}
}
\caption{(a) Temperature, integrated along the $z$ direction and
averaged with the electron density at $t$~= 14.51 min
in the MHD simulation.
Some magnetic field lines (green lines) and flux rope field lines (blue lines) are overplotted.
(b) Difference in the AIA observations in the 211~{\AA} passband
between 06:09:00.62 and 05:55:00.62 UT (the scale is in DN).}
\label{cfraia}
\end{figure}

Although the detailed comparison between the final state of the MHD simulation
and the corresponding state in the region under study poses
major challenges, it is still interesting to carry out a qualitative comparison
to have a global understanding of the weaknesses and strengths of the model.
Figure \ref{cfraia}a shows a map of the temperature averaged
for the electron density along the line of sight from the top view
at $t$~= 14.51 min in the MHD simulation.
Figure \ref{cfraia}b shows a difference image between the AIA 211~{\AA} pass band
at 06:09:00.62 and 05:55:00.62 UT.
The latter time is the closest AIA image to the assumed flux rope ejection time at 05:54:40 UT,
the former time is 14.3 minutes after, which is approximately the duration of the MHD simulation.
We find that the region in the center of the field of view,
where the emission in the AIA 211~{\AA} pass band is enhanced,
roughly corresponds to a region in the MHD simulation where the plasma is heated up on average (to about 4 MK)
and that a nearby location where the emission diminishes corresponds
to a relatively cold location in the MHD simulation.
In general, the MHD simulation shows an increase in temperature,
which is consistent with the generally enhanced emission in the 211~{\AA} pass band.
Also the magnetic field lines displayed in Figure \ref{cfraia}a describe
a topology with many similarities to the one suggested by the structures in Figure \ref{cfraia}b.
Examples include the system of loops in the left bottom corner of the field of view
and the expanding structures on the top right from the centre of the image.
They roughly correspond to the flux rope location at this time in the MHD simulation.

\begin{figure}   
\centerline{
\includegraphics[scale=0.40]{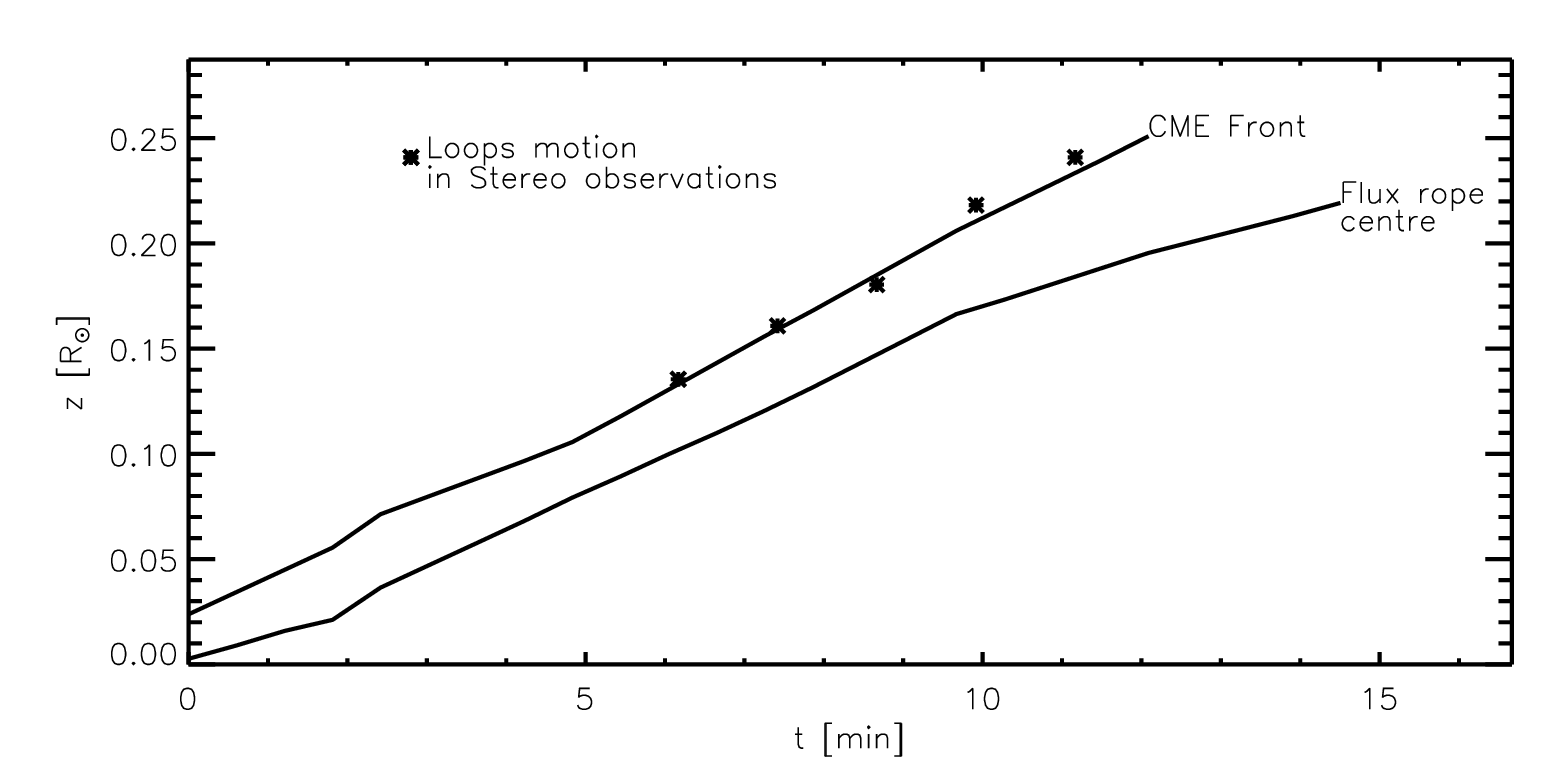}
}
\centerline{
\includegraphics[scale=0.40]{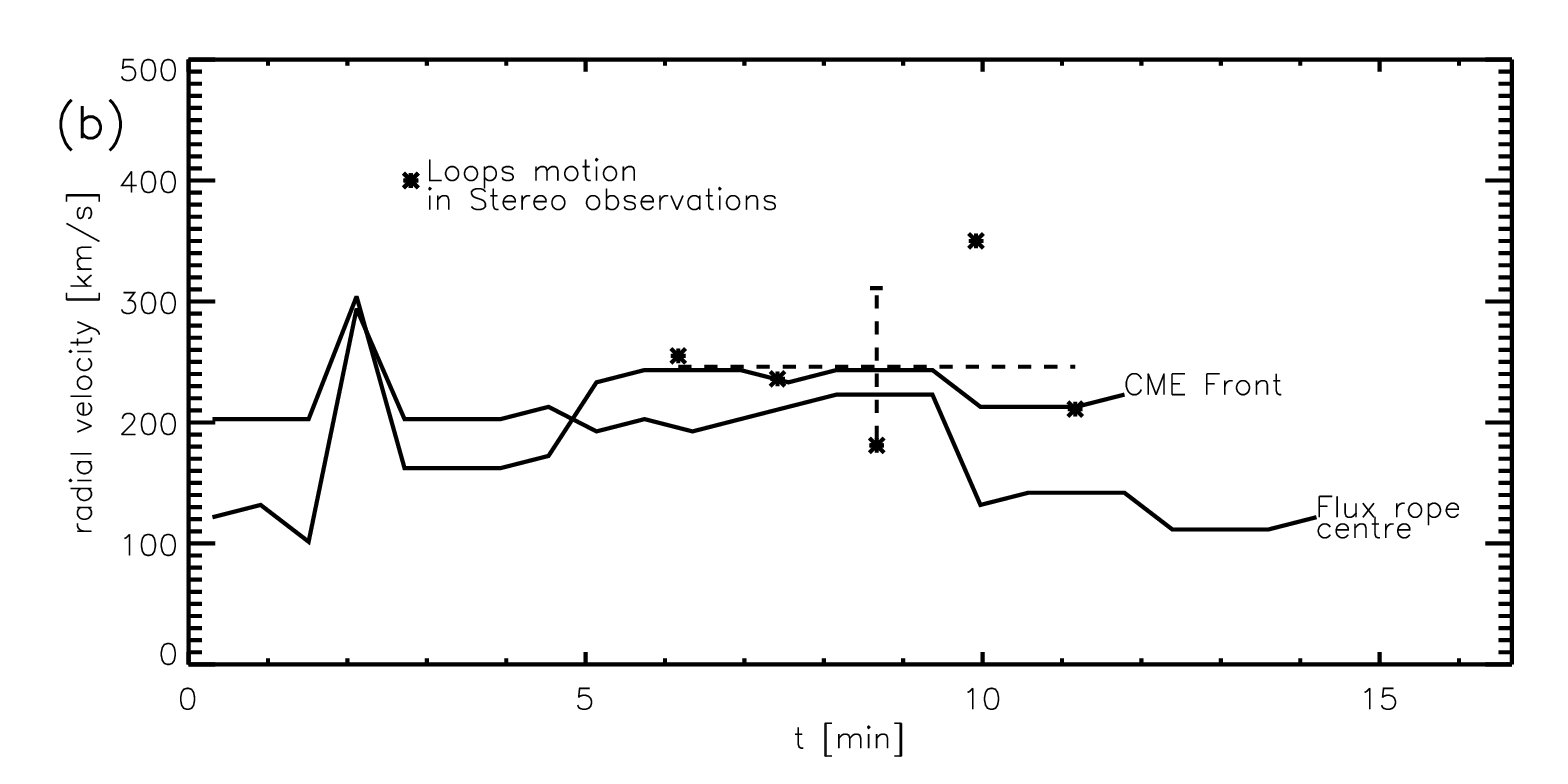}
}
\caption{(a) Comparison between the position of the centre of the flux rope,
the front of the ejecta and
the position of the upward expanding loops in STEREO images as a function of time.
(b) Comparison between the speed of the centre of the flux rope,
the speed of the front of the ejecta, and
the apparent speed of the expanding loops in STEREO images as a function of time.
Times are shown in minutes where 0 is the starting time of the MHD simulation at 06:00:41 UT.
Dashed lines represent the average speed of the expanding loops in STEREO observations with an error bar of $\pm$~65~km~s$^{-1}$.}
\label{frzv}
\end{figure}

Additionally, we have carried out a simple visual comparison of
the kinematics of the flux rope ejection with STEREO observations taken over a
time span after the observed start of the flux rope ejection (at 05:54:40 UT),
corresponding to the duration of the MHD simulation.
The visual tracking of the CME in the STEREO images lets us follow the apex of the expanding loops
that initiate the CME motion (cross points in Figure \ref{frzv}a).
In Figure 10a, these radial distances are compared with the tracked radial distance
of the centre of the flux rope and the front that propagates from its ejection in the MHD simulation.
We find a very good agreement of the locations identified in STEREO observations with the position of the flux rope front showing that probably the apex of the loops represents their motion as a consequence of being pushed by the ongoing ejection and represent the CME front.
The same good agreement is found in Figure \ref{frzv}b, where we
compare the speeds of propagation of the flux rope and the CME front
with the speed inferred from STEREO observations.
In the MHD simulation, the CME front accelerates from 100~km~s$^{-1}$ to about 200~km~s$^{-1}$,
while the flux rope moves at 200~km~s$^{-1}$ until the end of the simulation, where it slows down to 100~km~s$^{-1}$.
The points, tracked in STEREO images, always lie between 150\,--\,350~km~s$^{-1}$.
As we cannot consider this analysis further than a qualitative comparison, it is interesting to show that
the speed of the structure, predicted by the MHD simulation, is comparable to
the speed at which the structures move in the STEREO observations.
At the same time, it seems that the model underestimates the observed velocity
for at least a fraction of the time of evolution around $t$~= 200 s.
There, the observed speed is $\sim$~1.5 higher than the speed
of the flux rope centre and the front of the ejecta in the MHD model.
This difference would be enough to make the plasma flow overcome
the sound speed, thus leading to a shock in the MHD model matching the observation.
This may be a consequence of the differences between
the real atmospheric profile of $\rho$, $T$ compared to the inferred and simulated ones used here.

\section{Ion Charge State Evolution of CMEs} 
      \label{S-general}

The ion charge state of the CME plasma, as well as its evolution in the corona, depends mainly on the following factors: (i) the chemical element to which the considered ions belong, (ii) the plasma conditions (such as temperatures and electron densities), and (iii) the bulk velocities of plasma in the corona. In the previous sections we derived parameters of the erupting plasma and of the emerging flux rope at distances up to 0.25~R$_{\odot}$ from both direct EUV measurements and numerical simulations with the MHD model. In this section, we investigate the ion charge state evolution of the plasma structures under study by considering the ratios of carbon, C$^{6+}$/C$^{5+}$, and oxygen, O$^{7+}$/O$^{6+}$, and the average charge of iron ions, Q$_{\mathrm{Fe}}$, which were measured \textit{in situ} by ACE. For this purpose it is necessary to analyze how these parameters evolve during the plasma propagation in the corona at larger distances of several solar radii. We assume that in the whole space between the solar surface and the frozen-in region the expanding plasma is in a quasi-stationary state, \textit{i.e.} the plasma ionization and recombination time scales are less than the expansion time scale of the plasma.

Evolution of the ion charge states in the corona can be described by the following system of continuity equations for a set of ions from the atomic species of interest in the rest frame of the expanding plasma structure (see, \textit{e.g.} \citealp{Ko1997}):

   \begin{equation}
     \frac{\partial y_i}{\partial t} = N_e \left( y_{i-1}C_{i-1}(T_e) - y_i\left( C_{i}(T_e) + R_{i-1}(T_e)\right) + y_{i+1}R_{i}(T_e)  \right) \, ,
     \label{Eq-continuity}
   \end{equation}
where $y_i = n_i / \sum_{i=0}^Z n_i$ is the relative fraction of the ion with the number density $n_i$ in the charge state $i$, $N_e$ is the electron density, $T_e$ is the electron temperature, $C_i$ is the ionization coefficient rate for the transition from charge state $i$ to $i+1$, and $R_i$ is the total recombination rate (including both radiative and dielectronic recombination) from the charge state $i+1$ to $i$. For integrating the system of Equations (\ref{Eq-continuity}) we used the recombination and ionization rate coefficients $R_i$ and $C_i$ from the CHIANTI database (an atomic database for emission lines) \citep{Dere2007,Dere2009}, where these data are given on the assumption that the electron speed distribution is Maxwellian.

\begin{figure}  
\centering
\includegraphics[scale=0.6]{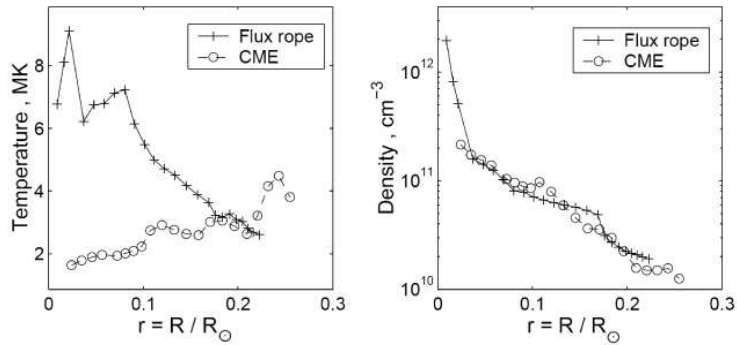}
\caption{Evolution of electron temperature and density of the flux rope and CME LE plasmas derived from the MHD simulations.}
\label{fig_fr_cme}
\end{figure}

In order to solve the system of equations, one needs to know the time evolution of the electron density and temperature, $T_e(t)$, $N_e(t)$, in the plasma structure under study as well as its bulk velocity, $v(t)$. As the plasma parameters, obtained from the EUV imaging and MHD modeling, are known only in the low corona up to the distance $r_0\approx$~0.25 (in units of R$_{\odot}$) from the solar surface, we model the evolution of plasma conditions at larger distances $r > r_0$ analytically, by taking into account the processes of cooling, heating, and expansion of the plasma, and solve the system of equations (Equations (\ref{Eq-continuity})) for the chosen ion species. As the initial conditions, we used the plasma parameters derived from the MHD simulation. In our analysis we separate two specific regions: the hot flux rope structure (hereafter referred to as “flux rope”) and the colder CME leading edge or compression front (hereafter referred to as “CME LE”, see Figure 5, \citealp{cheng2011}) surrounding the flux rope. The evolving profiles of electron temperatures and densities for the flux rope and the CME LE plasmas during the acceleration phase are presented in Figure \ref{fig_fr_cme}.

As seen in Figure \ref{fig_fr_cme}, the temperature of the flux rope rises up to 6--9 MK at the beginning of expansion and then drops down as it moves away, whereas the CME LE temperature evolution has the reverse trend. This result is consistent with the recent studies where hot flux rope structures were observed before and during the eruptive flare and CME events using SDO/AIA data (see \citealp{cheng2011,cheng2012,zhang2012,cheng2013}). In the work by \cite{nindos2015} almost half of the investigated eruptive events contained a hot flux rope configuration. The high flux rope temperatures in Figure \ref{fig_fr_cme} are inconsistent with those of the moving CME structure in Figure \ref{tem_den_evol}, because the latter corresponds to the colder outer shell of the CME (see Figures \ref{rundif}a, \ref{rundif}b). At the same time, the hot flux rope is not visible in Figure \ref{rundif}a due to the fact that the flare is brighter than its emission. This is in good agreement with the schematic model of the multi-temperature structure of the CME demonstrated by \cite{cheng2011} in their Figure 5.

The electron density at distances $r > r_0$ is taken to have a power law form

   \begin{equation}
      N_e(r) = N_e(r_0) \left(  \frac{r_0}{r}  \right)^3 \, ,
     \label{dens_distrib}
    \end{equation}
consistent with the flux rope evolution for expansion both in length and radius (see, \textit{e.g.} \cite{Kumar96,Lee09}). For the temperature profile we have used a form similar to the adiabatic relation

   \begin{equation}
      T_e(r) = T_e(r_0) \left(  \frac{N_e(r)}{N_e(r_0)}  \right)^\alpha \, ,
     \label{temp_distrib}
    \end{equation}
where the power index $\alpha$ is chosen to be consistent with the \textit{in situ} measured values of ion composition parameters. In the case of the adiabatic expansion the factor $\alpha = \gamma - 1$, where $\gamma =$ 5/3, is the adiabatic index. This simple form of the temperature profile was used to account for possible heating, which the ejected plasma usually undergoes long after the eruption (see, \textit{e.g.} \citealp{Akmal01,Ciaravella01}). Smaller values of the index $\gamma$ are often used in coronal models to take phenomenologically into account heating mechanisms without knowledge of heating rates (see also \citealp{Kumar96,Lee09}).

Using Equations (\ref{Eq-continuity}), we have carried out calculations of ion charge state evolution of C, O, and Fe ions for the event on 2 August, 2011. A relationship between cooling rates of plasma by different mechanisms, such as the adiabatic expansion, thermal conductive cooling, and radiative losses, depend on the plasma parameters and configuration of the erupting structure and varies during the expansion in the corona. We analyzed and estimated the cooling effectiveness of these three terms separately for our plasma conditions by considering the temperature evolution of the flux rope and the CME LE material. It was found that the most effective cooling factor is the adiabatic expansion. For instance, at the distance $r = 0.5$ the fall of temperature for the adiabatic regime prevails over those of the radiative and conductive cooling by 3--5 times, and at $r = 1$ it is up to one order of magnitude and more. Thus, it was shown that at distances $r>$~0.25 the adiabatic expansion is the noticeably prevalent the cooling process.

Assuming that the cooling is provided elsewhere only by the adiabatic expansion, we obtained the following frozen-in ion composition parameters: C$^{6+}$/C$^{5+}=$ 0.56, O$^{7+}$/O$^{6+}=$ 0.017, Q$_{\mathrm{Fe}}=$ 7.4 for the CME LE plasma, and C$^{6+}$/C$^{5+}=$ 0.071, O$^{7+}$/O$^{6+}=$ 1.46$\cdot$10$^{-4}$, Q$_{\mathrm{Fe}}=$ 7.4 for the flux rope structure. These values are too low in comparison with the \textit{in situ} observations (see Table \ref{T-ionstate}). If we then introduce a heating process in the region where the Fe ion state is frozen-in by decreasing the index $\alpha$ from the adiabatic value 2/3 to 0.1, the derived value of Q$_{\mathrm{Fe}}$ increases to $\sim~$10--11, which is in agreement with \textit{in situ} observations. However, in this case the frozen-in ratios C$^{6+}$/C$^{5+}$ and O$^{7+}$/O$^{6+}$ become noticeably larger by 2--3 times in comparison with the measurements.

In order to explain and to overcome this issue, we assumed that the heating power depends on the height in the corona. First, our numerical analysis showed that the ion charge states of C and O ions reach the frozen-in conditions at distances of about 1--2~R$_{\odot}$, whereas the frozen-in region of Fe ions begins at distances of $\approx$~4-5~R$_{\odot}$. The reason is that the recombination time scales of carbon and oxygen ions, $\tau_r = 1/(N_e R_i)$, prevail over those of the iron ions. Second, we considered separately two spatial intervals: the first one is from $r_0$ to $r_h =$ 1.5 where the frozen-in conditions begin to play a role only for C and O ions, and the second is from 1.5~R$_{\odot}$ to the Fe frozen-in region. We then calculated the evolution of the ion charge states of C and O ions by matching the parameter $\alpha$ so that to agree with the observational data in the first spatial interval. For the CME LE plasma we adopted the value $\alpha =$0.35 and found frozen-in values C$^{6+}$/C$^{5+}$~= 2.2 and O$^{7+}$/O$^{6+}$~= 0.32. For the flux rope structure $\alpha =$ 0.2, and C$^{6+}$/C$^{5+}$~= 1.77 and O$^{7+}$/O$^{6+}$~= 0.14. At the same time, at 1.5~R$_{\odot}$ we obtained Q$_{\mathrm{Fe}}\approx$~7--8, which is noticeably lower than the measured \textit{in situ} value.

Using the functional temperature difference between the proper adiabatic expansion and the fitted values of the index $\alpha$ for the CME LE plasma and flux rope, we also estimated the heating power from the coronal source, which maintains this difference. It was found that this source acted from $r_0$ to $r \approx$0.5 and its intensity sharply decreased with distance. The average heating power for both plasma structures are Q$_{\mathrm{CME}}\approx 5\cdot 10^{-3}$~erg~cm$^{-3}$~s$^{-1}$ for the CME LE plasma and Q$_{\mathrm{FR}}\approx 6\cdot 10^{-3}$~erg~cm$^{-3}$~s$^{-1}$ for the flux rope.

In order to match the value Q$_{\mathrm{Fe}}$ with the \textit{in situ} measurements, we assumed that in the second interval $r>$~1.5~R$_{\odot}$ the parameter Q$_{\mathrm{Fe}}$ increased due to an additional heating power from the coronal source. We have assumed a temperature evolution at distances $r_h \ge$1.5 as $T_e(r) = T_h (r/r_h)^\beta$. Taking the index $\beta =$ 0.75, we  obtained Q$_{\mathrm{Fe}}\approx$10 for the CME LE plasma and Q$_{\mathrm{Fe}}\approx$11 for the flux rope, which is compatible with the \textit{in situ} observations. Our estimation of the heating power at the point $r_h =$ 1.5 gives $Q_h \sim (1-2)\cdot 10^{-5}$~erg~cm$^{-3}$~s$^{-1}$.

\section{Discussion} 

In our study, we propose a method to predict the ion charge state of the ICME, produced by the flare and CME solar event on 2 August 2011, using the SDO/AIA EUV observations, MHD numerical simulation of the flux rope formation, and the analytical description of the plasma ion charge state evolution in the corona up to the frozen-in region. We assume that the ion composition of the ICMEs does not vary during their propagation in the heliosphere due to the very large ionization and recombination relaxation times at low plasma density in comparison with the travel time. In order to obtain the values of the ion charge state ratios C$^{6+}$/C$^{5+}$ and O$^{7+}$/O$^{6+}$ and the average charge Q$_{\mathrm{Fe}}$ close to those measured \textit{in situ}, we introduced a heating process, which had different rates at the distances 0.25--1.5~R$_{\odot}$ and 1.5--5~R$_{\odot}$.

The results of plasma diagnostics (see Section \ref{S-diagnostics}) and MHD simulations suggest that the erupting plasma is heated far from the flare region. Figure \ref{tem_den_evol} shows the temperature dropping to $\approx$~1.5~MK at the base of the ejected structure under consideration, whereas the MHD calculations exhibit noticeably higher temperatures for the hot flux rope and the colder CME LE plasmas. The assumption of a heating source above the flare region is of course debatable, but there is a number of works where the authors performed MHD simulations and discussed various heating mechanisms in the CME plasma (see, \textit{e.g.} \citealp{Lee09,Lynch11}). At the same time, the work of \citet{zhang2012} shows that in some cases the heating of the flux rope can start simultaneously with the ejection, as they observe the appearance of hot channel signatures at the earliest stage of the slow rise of the ejection of a magnetic flux rope.

\citet{Reinard12} carried out a numerical simulation of ion composition for two ICMEs (magnetic clouds), detected by STEREO and ACE on 21--23 May 2007, using the Magnetohydrodynamics-on-A-Sphere (MAS) and ARC7 ideal 2.5D MHD models, described earlier by \citet{Lynch11}. Both models took into account field-aligned thermal conduction, radiative losses, and coronal heating from the flare site during the initiation and expansion of a flux rope in the corona up to distances of more than 10~R$_{\odot}$. As it follows from their results, the key question in the prediction of the ICME ion composition by the numerical MHD modeling is a correct definition of the ratio between heating and cooling processes acting on the erupting plasma in the corona. In both models, described by \citet{Lynch11}, heating is introduced as the dominating factor not depending on real conditions in the source. Thus, they obtained that the slower CMEs became hotter than the faster ones, which contradicts the observations. On the contrary, we use a 3D MHD model to determine the plasma parameters of the ejecta on the initial stage of the flux rope ejection and fit the results with the EUV measurements by selecting the initial parameters of the simulation.

Our consideration refers to the case when the ICME parameters correspond to the apex of the CME. Though, in some cases the presence of the nearby coronal holes (CHs) producing high-speed streams can seriously influence the appearance and parameters of the ICME near Earth \citep{Gopals09a,Gopals09b,Mohamed12,Makela13,Wood12,Gopals13}. An interaction of the CME with a high-speed stream from the nearby coronal hole can deflect the CME from its initial direction, which results in the shifting of the arrival time ahead or behind the time, predicted by the kinematic models, and can change other solar wind parameters. We suppose that the ion composition parameters in the collisionless heliosphere are not influenced by such interaction and can be used for the source identification, but these cases are worth to be specially studied.

It should be noted, that our model is rather simplistic and idealized, as it does not reproduce some important details of the CME formation, such as laminar and turbulent features of the evolution visible in observations. Energy dissipation via electric resistivity, heat conduction, viscosity, and radiation should be especially taken into consideration during the early phase of the evolution in the dense corona. The used ideal one fluid MHD approach has a limited applicability in the case of the observed small scale and fast variations. Moreover, the numerical ideal MHD modelling, we have conducted in this work, does not take into account the terms in the energy equation that play a role in the coronal dynamics. Thermal conduction can be responsible of diffusing heat in the domain, even if \citet{pagano07} and \citet{pagano14} showed that it is largely inhibited inside the flux rope during the CME propagation. Dimensionless scaling and the relative importance of corresponding terms in the energy balance equations is not quite clear and needs more investigation for better understanding of the overall situation. Finally, a more accurate treatment for the effect of magnetic resistivity on the amount of magnetic energy converted into heating would require a much higher spatial resolution.

\section{Summary and Conclusion} 
      \label{S-Conclusion}
We presented a complex study of a series solar wind transients registered by ACE on 4 -- 7 August 2011 and their solar sources, flares and CMEs, which occurred on 2, 3 and 4 August 2011 in AR 11261. These events produced two shocks with sheaths and two ICMEs of the MC type, as identified by the RC list. The analysis of the ion charge state of the solar wind revealed three transients with enhanced temperature-dependent ratios C$^{6+}$/C$^{5+}$ , O$^{7+}$/O$^{6+}$ and a mean charge of iron ions Q$_{\mathrm{Fe}}$, which can be associated with hot plasma released in the coronal sources. The first transient, determined from the ion composition, coincided with the first ICME (Table 1), whereas the two others preceded the second ICME. The shift in time between transients 2, 3, and the second ICME may be probably caused by interaction between CME 2 and 3. Simulations with the WSA-Enlil cone model showed that the third CME of 4 August surpassed the second CME of 3 August at a distance from the Sun of about 0.6~AU.

We studied in detail the formation of the first CME of 2 August, using the SDO/AIA images in 211{\AA} and numerical simulations using NLFF and MHD modeling. The images of the eruptive structures in different SDO/AIA spectral channels were used for diagnostics of the outflow plasma by means of the DEM analysis. From the observational data it was found that in the event of 2 August the temperature of the plasma during its visible expansion from 0.1 to 0.13~R$_{\odot}$ decreased from 2.7 to 1.7~MK and the density from $1\cdot 10^9$ to $5\cdot 10^8$ cm$^{-3}$. These values are lower than those obtained by modeling for the apex, and correspond to the legs of the eruption shell. This confirms that heating is more effective in the upper part of the expanding structure.

The initiation of the CME of 2 August 2011 was simulated numerically, using a combination of the NLFF magnetic field extrapolation model with a 3D MHD model of the expanding flux rope, specially suited for the given case. The results of the simulation and comparison with the EUV measurements demonstrate that the general topology of the magnetic field matched the visible loop structure, whereas the flux rope was formed along the polarity inversion line and was pushed upwards by the unbalanced Lorentz force. The maximal speed was below the sound speed of 300~km~s$^{-1}$, thus, the model did not predict creation of a shock wave ahead of the flux rope ejection that was seen in the observations. The MHD simulation showed a temperature of $\sim$~4~MK in the CME apex, which coincides with an enhancement of radiation in the 211{\AA} channel. In the relative time scale starting at the moment of the flux rope raise-up, the simulated height time dependence of the CME structure up to the heights of 0.25~R$_{\odot}$ agrees well with the observations of STEREO in 171{\AA} at the limb, the difference in speed is within the measurement errors $\pm$~65~km~s$^{-1}$.

Based on the results of the observations and numerical simulation, the ion composition of CME1 in the frozen-in region in the event of 2 August 2011 was calculated with some assumptions about heating and cooling processes. The calculated values of the temperature-dependent ion ratios and the mean charge of iron agree with those measured \textit{in situ}, under the assumption that the expanding plasma was heated by an additional source. The average heating power decreased with height from $\sim (5-6)\cdot 10^{-3}$~erg cm$^{-3}$ s$^{-1}$ at $r_h \approx 0.5-1.5$ to $\sim (1-2)\cdot 10^{-5}$~[erg~cm$^{-3}$~s$^{-1}$] at $r_h \approx 1.5-5$.

In conclusion, our analysis of the ion composition of CMEs enables to disclose a relationship between parameters of solar wind transients and properties of their solar sources, which opens new possibilities to validate and improve the solar wind forecasting models.

This work was fulfilled as a contribution to the International Study of Earth-affecting Solar Transients (ISEST) Minimax 24 project (the event of 4 August 2011 is included into the ISEST event list\footnote{http://solar.gmu.edu/heliophysics/index.php/The\_ISEST\_Event\_List}).

\begin{acks}
 The authors are grateful to Jie Zhang and Nat Gopalswamy as the ISEST coordinators for supporting our studies of coronal sources of ICMEs. We thank Ian Richardson and Hilary Cane for their list of Near-Earth Interplanetary Coronal Mass Ejections \footnote{http://www.srl.caltech.edu/ACE/ASC/DATA/level3/icmetable2.htm}, CDAW Data Centre \footnote{http://cdaw.gsfc.nasa.gov/CME\underline{ }list/index.html}, and CACTus software package \footnote{http://sidc.oma.be/cactus/}, which we used in our investigations.
 The authors thank the GOES, SDO/AIA and ACE research teams for their open data policy. We are grateful for the opportunity to use the results of the simulation, obtained by  the WSA-Enlil Cone and DBM models\footnote{http://ccmc.gsfc.nasa.gov,http://helioweather.net/}. This project has received funding from the European Research Council (ERC) under the European Union's Horizon 2020 research and innovation programme (grant agreement No 647214). We acknowledge the use of the open source (gitorious.org/amrvac) MPI-AMRVAC software, relying on coding efforts from C. Xia, O. Porth, R. Keppens. The computational work for this paper was carried out on the joint STFC and SFC (SRIF) funded cluster at the University of St Andrews (Scotland, UK). The work is partially supported by RFBR grants 17-02-00787, 14-02-00945  and  the P7 Program of the Russian Academy of Sciences.
\end{acks}

\textbf{Disclosure of Potential Conflicts of Interest} The authors declare that they have no conflicts of interest.

The final publication is available at Springer via http://dx.doi.org/10.1007/s11207-017-1109-0.


\bibliographystyle{spr-mp-sola}
\bibliography{sola_bibliography}

\end{article}

\end{document}